\documentclass[usedcolumn,usenatbib,usegraphicx]{mn2e}

\title[Tidal Interaction in Coplanar Be/X-Ray Binaries] {Viscous
  Effects on the Interaction between the Coplanar Decretion Disc and
  the Neutron Star in Be/X-Ray Binaries} \author[A. T. Okazaki et al.]
{A. T.  Okazaki,$^{1, 2}$\thanks{E-mail:
    okazaki@elsa.hokkai-s-u.ac.jp} M.  R.  Bate,$^3$, G. I.
  Ogilvie$^2$,
  and J. E. Pringle,$^2$\\
  $^1$Faculty of Engineering, Hokkai-Gakuen University, Toyohira-ku,
  Sapporo 062-8605, Japan\\
  $^2$Institute of Astronomy, University of Cambridge,
  Madingley Road, Cambridge CB3 0HA\\
  $^3$School of Physics, University of Exeter, Stocker Road, Exeter
  EX4 4QL}

\date{Accepted.  Received; in original form}

\pagerange{\pageref{firstpage}--\pageref{lastpage}} \pubyear{2002}

\begin{document}

\maketitle

\label{firstpage}

\begin{abstract}
  We study the viscous effects on the interaction between the coplanar
  Be-star disc and the neutron star in Be/X-ray binaries, using a
  three-dimensional, smoothed particle hydrodynamics code. For
  simplicity, we assume the Be disc to be isothermal at the
  temperature of half the stellar effective temperature.  In order to
  mimic the gas ejection process from the Be star, we inject particles
  with the Keplerian rotation velocity at a radius just outside the
  star. Both Be star and neutron star are treated as point masses.  We
  find that the Be-star disc is effectively truncated if the
  Shakura-Sunyaev viscosity parameter $\alpha_{\rm SS} \ll 1$, which
  confirms the previous semi-analytical result.  In the truncated
  disc, the material decreted from the Be star accumulates, so that
  the disc becomes denser more rapidly than if around an isolated Be
  star. The resonant truncation of the Be disc results in a
  significant reduction of the amount of gas captured by the neutron
  star and a strong dependence of the mass capture rate on the orbital
  phase. We also find that an eccentric mode is excited in the Be disc
  through direct driving due to a one-armed bar potential of the
  binary.
  The strength of the mode becomes greater in the case of a smaller
  viscosity. In a high-resolution simulation with $\alpha_{\rm
  SS}=0.1$, the eccentric mode is found to precess in a prograde
  sense.  The mass capture rate by the neutron star modulates as the
  mode precesses.
\end{abstract}


\section{Introduction}
\label{sec:intro}

The Be/X-ray binaries represent the largest subclass of high-mass
X-ray binaries. About two-thirds of the identified systems fall into
this category. These systems consist of a Be star (i.e., a B star with
an equatorial disc) and, generally, a neutron star. The orbit is wide
(several tens of days $\la P_{\rm orb} \la$ several hundred days) and
eccentric ($0.1 \la e \la 0.9$).

Most of the Be/X-ray binaries show only transient X-ray activity due
to transient accretion of the circumstellar matter of the Be star,
while some show persistent X-ray emission. Each Be/X-ray binary
exhibits some or all of the following three types of X-ray activity:
\begin{enumerate}
\item periodic (Type~I) X-ray outbursts, coinciding with periastron
  passage ($L_{\rm X} \approx 10^{36-37}\,{\rm erg\, s}^{-1}$),
\item giant (Type~II) X-ray outbursts ($L_{\rm X} \ga 10^{37}\,{\rm
    erg\, s}^{-1}$), which show no clear orbital modulation,
\item persistent low-luminosity X-ray emission ($L_{\rm X} \la
  10^{34}\,{\rm erg\, s}^{-1}$)
\end{enumerate}
(\citealt*{swr86}; see also \citealt{neg98}).  These features imply a
complicated interaction between the Be-star envelope and the neutron
star.

A Be star has a two-component extended atmosphere, a polar region and
a cool ($\sim 10^4\,{\rm K}$) equatorial disc. The polar region
consists of a low-density, fast ($\sim 10^3\,{\rm km}\,{\rm s}^{-1}$)
outflow emitting UV radiation. The wind structure is well explained by
the so-called line-driven wind model, in which the radiative
acceleration results from the scattering of the stellar radiation in
an ensemble of spectral lines (\citealt*{cak75}; \citealt{abb82}). On
the other hand, the equatorial disc, which is geometrically thin and
nearly Keplerian, consists of a high-density plasma from which the
optical emission lines and the IR excess arise.  The radial velocity
of the disc is smaller than a few ${\rm km\,s}^{-1}$, at least within
$\sim 10$ stellar radii (\citealt{han94}; \citealt{han00};
\citealt{wm94}). Although there is no widely accepted model for discs
around Be stars, the viscous decretion disc model proposed by
\citet*{lso91} explains many of the observed features and thus seems
promising (\citealt{por99}; see also \citealt{oka01}). In this model,
the matter supplied from the equatorial surface of the star drifts
outwards because of the viscous effect and forms the disc. The basic
equations for viscous decretion discs are the same as those for
viscous accretion discs, except that the sign of $\dot{M}$ (mass
decretion/accretion rate) is opposite. The boundary conditions for
decretion discs, however, are different from those for accretion
discs. Therefore, the decretion disc has a structure different from
that of the accretion disc \citep{pri91}.

\begin{figure}
  \resizebox{\hsize}{!}{\includegraphics{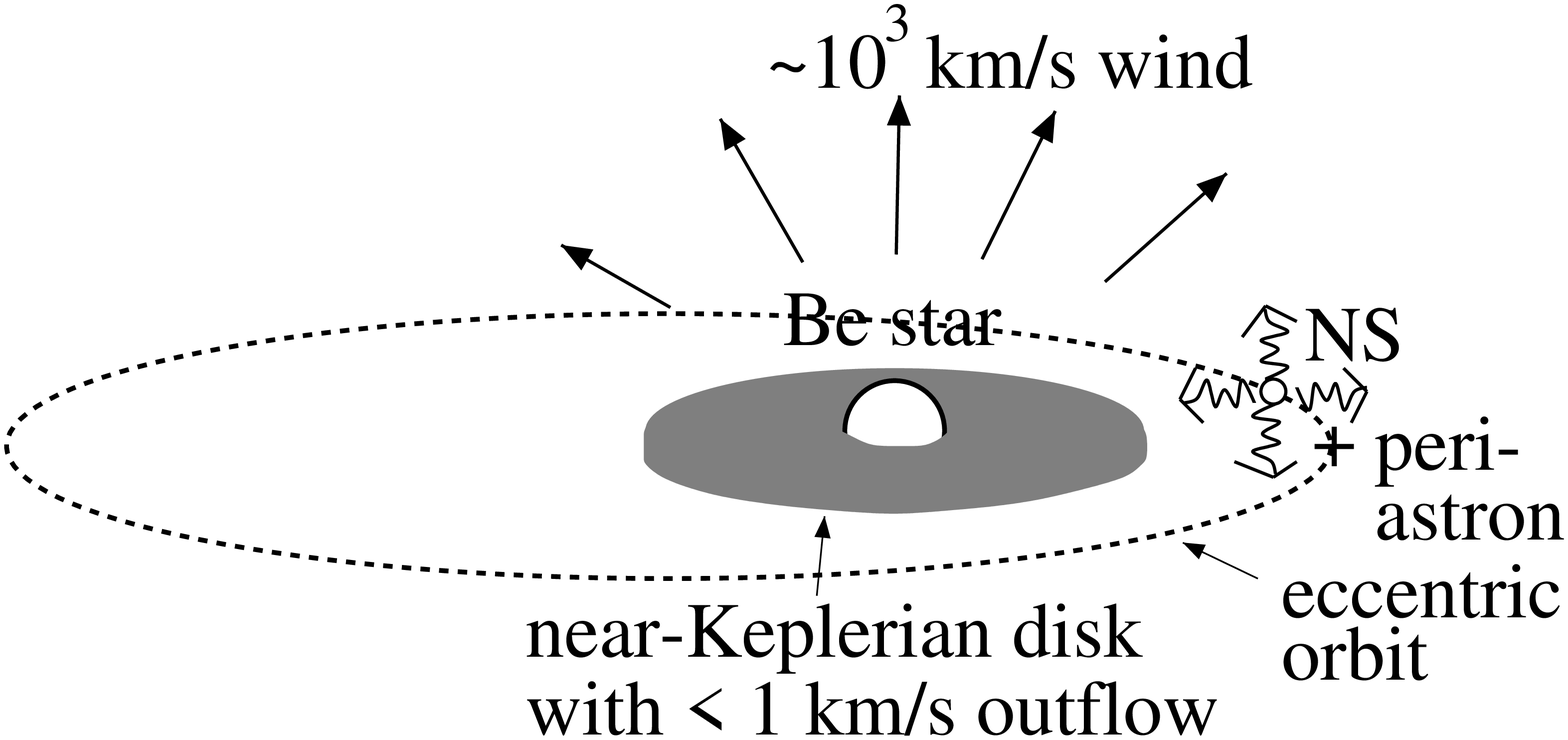}}
   \caption{Schematic view of a Be/X-ray binary,
     taken from \citet{on01b}.}
   \label{fig:bex}
\end{figure}

Until quite recently, models for Type~I X-ray outbursts in Be/X-ray
binaries had assumed a large disc around the Be star so that the
neutron star can accrete gas when it passes through the disc near
periastron. However, \citet{no01} and \citet{on01a} recently performed
a semi-analytical study based on the viscous decretion disc model for
Be stars and showed that the Be disc in Be/X-ray binaries is truncated
at a radius smaller than the periastron distance, as long as
$\alpha_{\rm SS} \ll 1$, where $\alpha_{\rm SS}$ is the
Shakura-Sunyaev viscosity parameter (see Fig.~\ref{fig:bex} for a
schematic view of a Be/X-ray binary). The truncation of the disc is
due to the resonant torque exerted by the neutron star, which removes
the angular momentum from the disc. The disc radii they obtained for
seven particular systems (4U\,0115+63, V\,0332+53, A\,0535+262,
EXO\,2030+375, 2S\,1417$-$624, GRO\,J1008$-$57, and 2S\,1417$-$624)
are consistent with the X-ray behaviour of those systems. Moreover,
the result is in agreement with the result of \citet*{rei97} that
there is a positive correlation between the orbital size and the
maximum equivalent width of H$\alpha$ ever observed in a system, a
measure of the maximum disc size around the Be star in the system.

The truncation of the Be disc in Be/X-ray binaries is not surprising.
The resonant interaction is important in various contexts even in a
fly-by encounter with a perturber. In fly-by encounters between a disc
galaxy and a point mass perturber, the energy is always transported
from the disc to the perturber through the resonant interaction,
except for overhead encounters where the energy transfer is small
\citep{pal83}. In distant encounters between a circumstellar accretion
disc and a perturbing mass with $r_{\rm peri}/r_{\rm disc} \ga 2$,
where $r_{\rm peri}$ and $r_{\rm disc}$ are the periastron distance
and disc radius, respectively, the disc material loses energy and
angular momentum to the perturber's orbit through a resonance feature
\citep*{hal96}.

In the case of Be/X-ray binaries, the surface density of the Be disc
is expected to increase more rapidly than that for isolated Be stars,
as a consequence of truncation. This qualitatively agrees with the
result found by \citet{zam01} that the Be discs in Be/X-ray binaries
are about twice as dense as those around isolated Be stars.  The disc
may finally become optically thick, and become unstable to
radiation-driven warping (\citealt{pri96}; see also \citealt{por98}).
Multi-wavelength, long-term monitoring observations of V635~Cas, the
optical counterpart of 4U\,0115+63, revealed that the Be disc in
4U\,0115+63 undergoes a quasi-cyclic ($3-5\,{\rm yr}$) dynamical
evolution \citep{neg01}: after each disc-loss episode, the disc starts
reforming, grows until it becomes unstable to warping, and after that
a Type~II outburst occurs. Although a direct link between the warped
disc and the Type~II outburst is still missing, the dynamical
evolution of the Be disc is likely to be the agent that controls the
X-ray behaviour of the system.

This way, the truncated disc model, at least qualitatively, explains
many of the observed features of Be/X-ray binaries. The
semi-analytical model adopted by \citet{no01} and \citet{on01a},
however, only compares the resonant torque integrated over the whole
orbit with the viscous torque to determine at which radius the disc is
truncated. Hence, it cannot make a quantitative prediction about how
perfect or imperfect the truncation is.  Moreover, it predicts nothing
about phase-dependent features, such as the disc deformation and the
change in the mass capture rate.

Therefore, in order to study the efficiency of the resonant truncation
and the orbital-phase dependence of the interaction, we simulate the
interaction between the Be-star disc and the neutron star in Be/X-ray
binaries, using a 3D SPH code. In a general context, such simulations
will also enable us to study the interaction between the viscous disc
and the companion in an eccentric orbit. In this paper, which is the
first of a series of papers dedicated to understanding the interaction
between the Be disc and the neutron star, we study the effects of
viscosity on disc truncation in a coplanar system.

\section{Calculations}
\label{sec:calculations}

\subsection{SPH Code}
\label{sec:sph_code}

Simulations presented here were performed with a three-dimensional,
smoothed particle hydrodynamics (SPH) code. The SPH code is based on a
version originally developed by Benz (\citealt{ben90};
\citealt{bbcp90}). The smoothing length is variable in time and space.
The code uses a tree structure to calculate the nearest neighbours of
particles. The SPH equations with the standard cubic spline kernel are
integrated using a second-order Runge-Kutta-Fehlberg integrator with
individual time steps for each particle \citep*{bat95}, which results
in an enormous computational saving when a large range of dynamical
time-scales are involved.

In our code, the Be disc is modelled by an ensemble of gas particles,
each of which has a negligible mass chosen to be $10^{-10} M_{\sun}$
with a variable smoothing length.  For simplicity, the gas particles
are assumed to be isothermal at the temperature of $T_{\rm eff}/2$,
where $T_{\rm eff}$ is the effective temperature of the Be star.  On
the other hand, the Be star and the neutron star are modelled by two
sink particles \citep{bat95} with corresponding masses. Gas particles
which fall within a specified accretion radius are accreted by the
sink particle.  We assume that the Be star has the accretion radius of
$0.9 R_{*}$, $R_{*}$ being the radius of the Be star.  For the neutron
star, we adopt a variable accretion radius of $0.9 r_{\rm L}$, where
$r_{\rm L}$ is the Roche-lobe radius for a circular binary. This is
because having a small accretion radius is time-consuming and it is
unphysical to adopt an accretion radius smaller than the smoothing
length of the particles near the disc outer radius.  An approximate
formula for $r_{\rm L}$ is given by
\begin{equation}
   r_{\rm L} \simeq 0.462 \left( \frac{q}{1+q} \right)^{1/3} D
   \label{eq:roche}
\end{equation}
(e.g., \citealt{war95}) with the mass ratio $q=M_{X}/M_{*}$, where
$M_{X}$ and $M_{*}$ are the masses of neutron star and Be star,
respectively, and the distance between the stars, $D$.

The SPH artificial viscosity, $\Pi_{ij}$, between particles $i$ and
$j$ enters the momentum equation as
\begin{equation}
   \frac{d \bmath{v}_i}{d t} =
   -\sum_j m_j \left( \frac{P_i}{\rho_i^2} + \frac{P_j}{\rho_j^2}
               +\Pi_{ij} \right) \nabla_i W(r_{ij}, h_{ij}),
   \label{eq:mom}
\end{equation}
where $\bmath{v}$ is the velocity, $m$ is the mass, $P$ is the
pressure, $\rho$ is the density, $W$ is the standard cubic spline
kernel, $r_{ij}$ is the distance between particles $i$ and $j$,
$h_{ij}=(h_i+h_j)/2$ is the mean of the smoothing lengths of particles
$i$ and $j$, and $\Pi_{ij}$ has the following standard form,
\begin{equation}
   \Pi_{ij} = \left\{
               \begin{array}{ll}
                  \left( -\alpha_{\rm SPH} c_{\rm s} \mu_{ij}
                   +\beta_{\rm SPH}\mu_{ij}^{2}\right)/\rho_{ij}
                  & \bmath{v}_{ij}\bld{\cdot}\bmath{r}_{ij} \le 0\\
                  0 & \bmath{v}_{ij}\bld{\cdot}\bmath{r}_{ij} > 0,
               \end{array} \right.
   \label{eq:Pi}
\end{equation}
\citep{mg83}, where $\alpha_{\rm SPH}$ and $\beta_{\rm SPH}$ are the
linear and nonlinear artificial viscosity parameters, respectively,
$\rho_{ij}=(\rho_i+\rho_j)/2$,
$\bmath{v}_{ij}=\bmath{v}_i-\bmath{v}_j$, $c_{\rm s}$ is the
isothermal sound speed, and
$\mu_{ij}=h_{ij}\bmath{v}_{ij}\bld{\cdot}\bmath{r}_{ij}/(r_{ij}^2+\eta^2)$
with $\eta^2=0.01h^2$.

In the continuum limit, the viscous force $\bmath{F}_{\rm v}$ in the
above SPH formalism is written as
\begin{equation}
   \bmath{F}_{\rm v} = \frac{\alpha_{\rm SPH} h \kappa}{2 \rho}
        \left[ \bld{\nabla \cdot}(c_{\rm s} \rho \bmath{S})
        +\nabla (c_{\rm s} \rho \bld{\nabla \cdot}\bmath{v})\right]
   \label{eq:fv1}
\end{equation}
\citep*{meg93}, where
\begin{equation}
   \kappa = -\frac{4\pi}{15} \int r^3 \frac{dW}{dr} dr
          = \frac{1}{5}
   \label{eq:kappa}
\end{equation}
in the 3D SPH code with the cubic spline kernel and
\begin{equation}
   S_{ij} = \frac{\partial v_i}{\partial j}
            +\frac{\partial v_j}{\partial i}
\end{equation}
is the deformation tensor.  If we assume that the density varies on a
length-scale much longer than the velocity, we have
\begin{equation}
   \bmath{F}_{\rm v} = \frac{1}{10} \alpha_{\rm SPH} c_{\rm s} h
                       \left[ \nabla^2 \bmath{v} + 2 \nabla (\nabla
                       \cdot \bmath{v}) \right].
   \label{eq:fv2}
\end{equation}
This implies that the shear viscosity $\nu$ and the bulk viscosity
$\nu_{\rm b}$ are given by
\begin{equation}
   \nu = \frac{1}{10} \alpha_{\rm SPH} c_{\rm s} h
   \label{eq:nusph}
\end{equation}
and
\begin{equation}
   \nu_{\rm b} = \frac{5}{3} \nu,
   \label{eq:nubsph}
\end{equation}
respectively, in the continuum limit of the 3D SPH formalism.  On the
other hand, in the Shakura-Sunyaev viscosity prescription, the shear
viscosity $\nu$ is written as
\begin{equation}
   \nu = \alpha_{\rm SS} c_{\rm s} H,
   \label{eq:nuss}
\end{equation}
where $H$ is the scale-height of the disc. From equations~(\ref{eq:nusph})
and (\ref{eq:nuss}), we have the relation between
the SPH artificial viscosity parameter $\alpha_{\rm SPH}$ and the
Shakura-Sunyaev viscosity parameter $\alpha_{\rm SS}$ as 
\begin{equation}
   \alpha_{\rm SS} = \frac{1}{10} \alpha_{\rm SPH} \frac{h}{H},
   \label{eq:alphass}
\end{equation}
if we assume that $\bld{\nabla \cdot}\bmath{v}=0$.
In general flows, however, $\bld{\nabla \cdot}\bmath{v} \ne
0$. Moreover, the viscosity is artificially turned off for divergent
flows in our model [see equation~(\ref{eq:Pi})]. Therefore,
equation~(\ref{eq:alphass}) should be taken as a rough approximation to the
relation between  $\alpha_{\rm SS}$ and $\alpha_{\rm SPH}$.

In several simulations we report in this paper, we adopted
$\alpha_{\rm SPH}=1$ and $\beta_{\rm SPH}=2$, in which case $\alpha_{\rm
  SS}$ is variable in time and space.  In the other simulations,
however, we adopted constant values of $\alpha_{\rm SS}$ in order to
roughly model the $\alpha$ disc. In these simulations, $\alpha_{\rm
  SPH}=10 \alpha_{\rm SS} H/h$ was variable in time and space and
$\beta_{\rm SPH}=0$.

The mass ejection mechanism from the Be star is not known. In our
simulations, it is modelled by injecting gas particles at $r = r_{\rm
 inj}$. In our normal resolution simulations, which finally have $\sim
20,000$ particles, we take $r_{\rm inj} = 1.02 R_{*}$, while in a
high-resolution simulation we performed with $\alpha_{\rm SS}=0.1$,
which finally has about $140,000$ particles, we take $r_{\rm inj} =
1.01 R_{*}$. These values were adopted so that $r_{\rm inj} \sim R_{*}
+ \bar{h}/2$, where $\bar{h}$ is a typical smoothing length near the
stellar surface. In order to avoid an additional complexity, we kept
the injection rate constant in each simulation.  Once injected, gas
particles interact with each other. As a result, most of the injected
particles fall onto the Be star by losing the angular momentum and a
small fraction of particles drift outwards, getting the angular
momentum from the other particles.

As the Be star, we take a B0V star with $M_{*} = 18 M_{\sun}$, $R_{*}
= 8 R_{\sun}$, and $T_{\rm eff} = 26,000\,{\rm K}$, which has the
typical spectral type of Be/X-ray binaries.  With these
  parameters, the disc scale-height $H$ is $\sim 0.02r$ at $r=R_{*}$
  and increases as $r^{3/2}$. For the neutron star, we take
$M_X=1.4M_{\sun}$ and $R_X=10^6\,{\rm cm}$.

\subsection{Initial Configuration}
\label{sec:config_ini}

We set the binary orbit on the $x$-$z$ plane with the major axis
along the $x$-axis.  At $t=0$, the neutron star is at apastron.
It orbits about the Be star primary with the orbital period $P_{\rm
  orb}$ and the eccentricity $e$. The angle of misalignment, $\beta$,
between the binary orbital plane and the Be disc plane is set to 0 in
this paper.  The unit of time is $P_{\rm orb}$, unless noted
otherwise.

Each simulation is very time-consuming. It takes three to four weeks
to perform each of the normal-resolution simulations, which runs on a
single processor of an HP Exemplar V2500, and about four months to
perform the high-resolution simulation, which runs efficiently on
eight processors of an SGI Origin 3800.  This is particularly so
for long-period systems. Thus, in this paper, we present results for
only a restricted range of parameter space as the first step in the
study of the interaction between the Be disc and the neutron star. We
fix the orbital period $P_{\rm orb}$, the orbital eccentricity $e$,
and the misalignment angle $\beta (=0)$, allowing only the viscosity
parameter to vary.  We adopt $P_{\rm orb}=24.3{\,d}$ and $e=0.34$,
targeting 4U\,0115+63, one of the best studied Be/X-ray binaries.
Then, the semi-major axis $a = 6.6 \cdot 10^{12}\,{\rm cm} \sim 12
R_{*}$ and $r_{\rm L}$ given by equation~(\ref{eq:roche}) ranges
between $0.13 a$ at periastron and $0.26 a$ at apastron.

\subsection{Testing the Code: Discs around Isolated Be Stars}
\label{sec:test}

Before attempting the binary simulations, we applied the code to an
isolated Be star in order to test whether it gives reliable results
compared to those obtained by using a one-dimensional
diffusion-type equation for the surface density.

\subsubsection{One-Dimensional Model}
\label{sec:1d_single}

As mentioned above, we adopt the viscous decretion disc model, which
yields a geometrically thin, nearly Keplerian disc around the Be star.
For simplicity, we assume that the disc around an isolated Be star is
axisymmetric and Keplerian and in hydrostatic equilibrium in the
vertical ($z$-) direction.  The evolution of such a disc is described
by
\begin{equation}
   \frac{\partial \Sigma}{\partial t} = \frac{1}{r}
   \frac{\partial}{\partial r} \left[
   \frac{\displaystyle \frac{\partial}{\partial r} \left(
         \alpha_{\rm SS} c_{\rm s}^2 r^2 \Sigma \right)}
        {\displaystyle \frac{d}{dr} \left( r^2 \Omega \right)}
   \right]
   \label{eq:sigma1d}
\end{equation}
with
\begin{equation}
   V_{\rm r}=
        -\frac{\displaystyle \frac{\partial}{\partial r} \left(
         \alpha_{\rm SS} c_{\rm s}^2 r^2 \Sigma \right)}
        {\displaystyle r \Sigma \frac{d}{dr} \left( r^2 \Omega \right)}
   \label{eq:vr1d}
\end{equation}
(e.g., \citealt{pri81}), where $\Sigma$ is the surface density,
$V_{\rm r}$ is the radial velocity, and $\Omega = (GM_{*}/r^3)^{1/2}$
is the angular frequency of disc rotation.

In order to create a situation similar to that in SPH simulations, we
inject mass at a constant rate at $r_{\rm inj} = 1.02 R_{*}$. At the
inner boundary $r=R_{*}$, we adopt the torque-free boundary condition,
$\Sigma = 0$.  We also take $\Sigma = 0$ as the outer boundary
  condition at $r = 10^3R_{*}$, which affects the disc structure only
  in a region near $r = 10^3R_{*}$.

\begin{figure}
  \resizebox{\hsize}{!}{\includegraphics{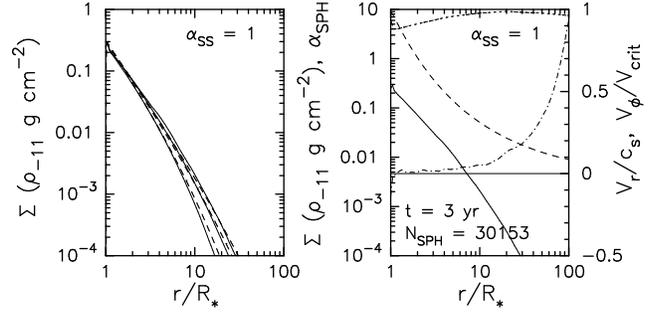}}
   \caption{Evolution of the viscous decretion disc around an
     isolated Be star in a 3D SPH simulation with $\alpha_{\rm SS}=1$.
     In the left panel, the surface density evolution for the first
     three years is shown by the solid lines ($t =$ 1\,yr, 2\,yr and
     3\,yr from left). The surface density is measured in units of
     $\rho_{-11}{\rm g}\,{\rm cm}^{-2}$, where $\rho_{-11}$ is the
     highest local density at $t= 1$\,yr normalised by $10^{-11}{\rm
     g}\,{\rm cm}^{-3}$, a typical value for Be stars. For comparison,
     the surface density distribution at the same epochs in a
     corresponding 1D model is shown by the dashed lines. The right
     panel shows the disc 
     structure at $t=3$\,yr.  The solid, the dashed, the dash-dotted,
     and the dotted lines denote the surface density, the azimuthal
     velocity normalised by the critical velocity of the Be star, the
     radial Mach number, and $\alpha_{\rm SPH}$. In both panels, the
     density is integrated vertically and averaged azimuthally, while
     the velocity components and $\alpha_{\rm SPH}$ are averaged
     vertically and azimuthally. The profile of $V_{\phi}$ is
     indistinguishable from one proportional to $r^{-1/2}$.
     Annotated at the bottom of the right panel is the number
     of SPH particles at this epoch.}
   \label{fig:isolated_a1}
\end{figure}

The evolution of $\Sigma$ for $\alpha_{\rm SS}=1$ and
$\alpha_{\rm SS}=0.1$ for the initial three years is shown by
dashed lines ($t=$ 1\,yr, 2\,yr and 3\,yr from left) in the left
panels of Figs.~\ref{fig:isolated_a1} and \ref{fig:isolated_a01},
respectively. 
The surface density is measured in units of $\rho_{-11}{\rm g}\,{\rm
  cm}^{-2}$, where $\rho_{-11}$ is the highest local density at $t=
1$\,yr normalised by $10^{-11}{\rm g}\,{\rm cm}^{-3}$, a typical value
for Be stars \citep{war88}. It should be noted that, in this
framework, no steady solution is present for decretion discs unlike
for accretion discs \citep{pri91}. Instead, the formal solution of
equation~(\ref{eq:sigma1d}) with $\partial \Sigma/\partial t = 0$ and
$V_{\rm r}=0$ gives the disc structure at $t \to \infty$, which is
given by $\Sigma \sim r^{-2}$ in our isothermal disc model.

\subsubsection{3D SPH Simulations}
\label{sec:sph_single}

Using the model described in $\S$\ref{sec:sph_code}, we performed two
3D SPH simulations of the disc evolution around an isolated Be star
with $\alpha_{\rm SS}=1$ and $\alpha_{\rm SS}=0.1$. In these
simulations, $\alpha_{\rm SPH}$ is variable in time and space and
$\beta_{\rm SPH}=0$ to keep $\alpha_{\rm SS}$ constant, as described
in the previous section.

\begin{figure}
  \resizebox{\hsize}{!}{\includegraphics{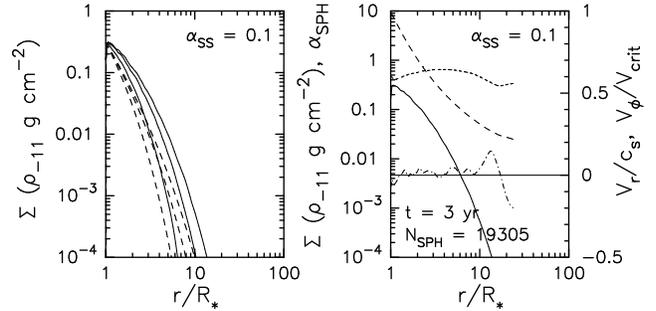}}
   \caption{Same as Fig.~\ref{fig:isolated_a1}, but for
     $\alpha_{\rm SS}=0.1$.}
   \label{fig:isolated_a01}
\end{figure}

Figs.~\ref{fig:isolated_a1} and \ref{fig:isolated_a01} show
the surface density evolution (left) and the disc structure at
$t=3$\,yr (right) for $\alpha_{\rm SS}=1$ and $\alpha_{\rm SS}=0.1$,
respectively. The time interval between adjacent contours in the left
panel is 1\,yr. In the right panel, the solid, the dashed, and the
dash-dotted lines denote the surface density, the azimuthal velocity
normalised by the stellar critical velocity, and the radial Mach
number, respectively. In the figures, the density is integrated
vertically and averaged azimuthally, while the velocity components are
averaged vertically and azimuthally. The dotted line shows the
vertically and azimuthally averaged distribution of $\alpha_{\rm
  SPH}$ required to keep $\alpha_{\rm SS}$ constant.

From Figs.~\ref{fig:isolated_a1} and \ref{fig:isolated_a01},
we observe that our 3D SPH code is capable
of producing the disc evolution similar to that in the 1D model,
except that for small viscosity the disc evolves a bit faster than in
the 1D model. The disc structure is almost Keplerian and the radial
velocity, which decreases with time, is very subsonic for $r \la
10R_{*}$ at $t \ga 1$\,yr.  These features are in agreement with the
observed characteristics of Be stars.

\begin{figure}
  \resizebox{\hsize}{!}{\includegraphics{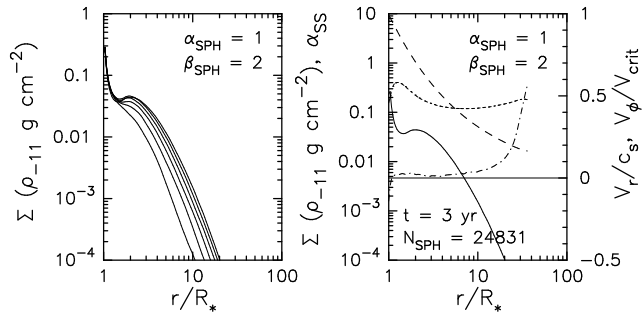}}
  \caption{Same as Fig.~\ref{fig:isolated_a1}, but for
    $\alpha_{\rm SPH}=1$ and $\beta_{\rm SPH}=2$. The dotted line in
    the right panel denotes the distribution of $\alpha_{\rm SS}$,
    after being averaged vertically and azimuthally.}
   \label{fig:isolated_sph2}
\end{figure}

For the purpose of comparison, we present in Fig.~\ref{fig:isolated_sph2} the
result from a simulation with $\alpha_{\rm SPH}=1$ and $\beta_{\rm
  SPH}=2$. In this simulation, $\alpha_{\rm SS}$ is variable in time
and space and is roughly proportional to $\rho^{-1/3} r^{-3/2}$. In
the initial phase of disc formation, the density distribution has a
very steep slope in the radial direction, so that $\alpha_{\rm SS}$
stays high except for a region near the injection radius. As time goes
on, $\alpha_{\rm SS}$ in the outer region decreases, because of the
increase in the density there. As a result, $\alpha_{\rm SS}$ has a
local maximum near the star, which, in turn, causes a dip in the
density distribution, the feature not seen in simulations with
constant $\alpha_{\rm SS}$.

Fig.~\ref{fig:isolated_sph2} also shows that the disc structure
outside the dip ($r \ga 3 R_{*}$) is, in a rough sense, similar to
that in simulations with constant $\alpha_{\rm SS}$. Since our purpose
is to study the interaction between the Be disc and the neutron star
in a system, of which the periastron distance is $\sim 8 R_{*}$, the
difference in the disc structure close to the star does not matter
very much. In the following sections, we will see that both kinds of
models give similar results.

\begin{figure}
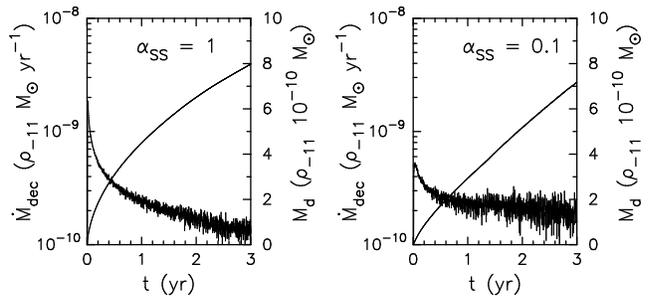

  \resizebox{\hsize}{!}{\includegraphics{fig5a.eps}
    \quad\quad \includegraphics{fig5b.eps}}
   \caption{Evolution of the mass-loss rate from the star,
    $\dot{M}_{\rm dec}$, (thin line) and the disc mass $M_{\rm disc}$
     (thick line) for $\alpha=1$ (left) and $\alpha=0.1$ (right). The
     mass-loss rate is measured at $r = R_{*}+2r_{\rm inj}$ and
     averaged over $\sim 9$ days.}
   \label{fig:isosph_mdot}
\end{figure}

In our model, in which the mass injection rate is kept constant, the
net mass-loss rate from the star through the disc decreases as the
disc mass increases. This is because a larger fraction of injected
particles must lose the angular momentum and fall back to the star to
support a larger disc. 
Fig.~\ref{fig:isosph_mdot} gives the change in the mass-loss rate from
the star, $\dot{M}_{\rm dec}$, (thin line) measured at $r = 2r_{\rm
  inj}-R_{*}$ and the disc mass $M_{\rm d}$ (thick line) for the
simulations shown in Figs.~\ref{fig:isolated_a1} and
\ref{fig:isolated_a01}. To reduce the fluctuation noise,
$\dot{M}_{\rm dec}$ is averaged over $\sim 9$ days. As expected, a
larger viscosity makes the decrease in $\dot{M}_{\rm dec}$ faster.
From the figure, we note that, for the first several years,
$\dot{M}_{\rm dec} \sim$ several $\times 10^{-10} \rho_{-11}
M_{\sun}\, {\rm yr}^{-1}$ for a wide range of the viscosity parameter.

It is interesting to compare the model mass-loss rate with the
observed one. The observed equatorial mass-loss rate is, however, a
poorly determined quantity. It has been measured only during the
disc-formation phase for several stars. Among them, X~Per is the only
one star for which both the equatorial mass-loss rate and the highest
disc density are known. \citet{tel98} studied the long-term behaviour
of the Be disc of X~Per (4U\,0352+30), a Be/X-ray binary system with
$P_{\rm orb}=250\,{\rm d}$ and $e=0.11$, and found that the equatorial
mass-loss rate for a disc build-up phase of less than 230\,d was
greater than $5.3 \times 10^{-9} M_{\sun}\,{\rm yr}^{-1}$ and that for
the following 380\,d was $3.7 \times 10^{-9} M_{\sun}\,{\rm yr}^{-1}$.
They also found that the base density of the Be disc of X~Per was as
high as $1.5 \times 10^{-10}\,{\rm cm}^{-3}$.  It should be noted that
the model mass-loss rate shown in Fig.~\ref{fig:isosph_mdot} is in
good agreement with the observed equatorial mass-loss rate of X~Per.

The simulations shown above assume a continuous mass supply from the
Be star for studying the formation and evolution of a persistent
disc. Some Be stars exhibit a transient disc formation by an episodic
mass loss. Such a situation was investigated by \citet{kh97}. They
studied the evolution of the gas explosively ejected from a Be star to
model the transient disc formation and disc decay, using a 3D SPH
code. They found that the gaseous particles form a nearly Keplerian disc
in the viscous time-scale. Because of the difference between the
boundary conditions adopted in this paper and by \citet{kh97}, the
structure of our persistent disc is different from that of the
decaying disc studied by \citet{kh97}.

\section{B\lowercase{e} Disc-Neutron Star Interaction in a Coplanar
  System}
\label{sec:sph_binary}

In the previous section, we have seen that our model with a 3D SPH
code is capable of reproducing the results from 1D simulations of
formation and evolution of the disc around isolated Be stars and
explaining the observed equatorial mass-loss rate from the Be star.
In this section, we apply our model to a Be/X-ray binary with $P_{\rm
  orb}=24.3\,{\rm d}$ and $e=0.34$, assuming coplanarity between
the Be disc and the binary.  The semi-major axis of the binary,
  $a$, is $6.6 \times 10^{12}\,{\rm cm}$ ($\sim 12R_{*}$). We have
run simulations with $\alpha_{\rm SS}=1$ for $30P_{\rm orb}$ ($\sim
2.0\,{\rm yr}$), by which time the disc almost reaches an equilibrium
state, while we have run the other simulations for $50P_{\rm orb}$
($\sim 3.3\,{\rm yr}$). In all simulations but one, the number of SPH
particles at the end of the simulation was about $2 \times 10^{4}$.
In order to confirm the reliability of the results obtained by those
simulations, we performed a simulation with $\alpha_{\rm SS}=0.1$ with
a higher resolution, in which the number of SPH particles at the end
of the simulation was about $1.4 \times 10^{5}$.

\begin{table*}
   \begin{minipage}{170mm}
      \caption{Summary of the results from binary simulations.
        The system has $P_{\rm orb}=24.3\,{\rm d}$ and $e=0.34$ in all
        simulations.  \lq var' in the viscosity-parameter columns
        means that the quantity is variable in time and space. For the
        $\alpha_{\rm SS}=1$ simulation, $r_{\rm d}$ is the radius at
        which the surface density distribution has a break, while for
        the other simulations it is the radius at which the disc is
        truncated.  $\dot{M}_{\rm dec}$ and $\dot{M}_{\rm acc}$ are
        the net mass-loss rate from the Be star and the mass capture rate
        by the neutron star, respectively, in units of $\rho_{-11}
        M_{\sun}\,{\rm yr}^{-1}$, where $\rho_{-11}$ is the highest
        local density at $t= 1$\,yr normalised by $10^{-11}{\rm
          g}\,{\rm cm}^{-3}$, and $S_{1,0}$ is the strength of the
        (1,0) mode. $<\cdots>$ denotes the average over $t_{\rm
        f}-5P_{\rm orb} \le t \le t_{\rm f}$. \lq ---' in
        the last column means that no precession is seen.}
      \begin{tabular}{ccccrrccccc}
         \hline
         \multicolumn{3}{c}{Viscosity parameters} &
         Run time &
         \multicolumn{2}{c}{$N_{\rm SPH}$} &
         \multicolumn{1}{c}{$<r_{\rm d}>$} &
         \multicolumn{1}{c}{$<\dot{M}_{\rm dec}>$} &
         \multicolumn{1}{c}{$<\dot{M}_{\rm acc}>$} &
         \multicolumn{2}{c}{Eccentric mode} \\
         $\alpha_{\rm SS}$ & $\alpha_{\rm SPH}$ & $\beta_{\rm SPH}$ &
         $t_{\rm f}$ ($P_{\rm orb}$) &
         \multicolumn{1}{c}{initial} &
         \multicolumn{1}{c}{final} &
         \multicolumn{1}{c}{($a$)} &
         \multicolumn{1}{c}{($\rho_{-11} M_{\sun}\,{\rm yr}^{-1}$)} &
         \multicolumn{1}{c}{($\rho_{-11} M_{\sun}\,{\rm yr}^{-1}$)} &
         \multicolumn{1}{c}{$<S_{1,0}>$} &
         \multicolumn{1}{c}{$P_{\rm prec}$ ($P_{\rm orb}$)} \\
         \hline
         1 & var & 0 & 30 & 100 & 16227 & 0.39 &
             $2.4\,10^{-10}$ & $2.1\,10^{-10}$ &
             $3.7\,10^{-2}$ & --- \\
         0.3 & var & 0 & 50 & 100 & 20820 & 0.38 &
             $1.7\,10^{-10}$ & $9.9\,10^{-11}$ &
             $8.9\,10^{-2}$ & --- \\
         var & 1 & 2 & 50 & 600 & 19661 & 0.37 &
             $3.0\,10^{-10}$ & $1.7\,10^{-10}$ &
             $9.6\,10^{-2}$ & --- \\
         0.1 & var & 0 & 50 & 100 & 19232 & 0.37 &
             $1.9\,10^{-10}$ & $3.9\, 10^{-11}$ &
             $1.0\,10^{-1}$ & --- \\
         0.1 & var & 0 & 47 & 1000 & 140108 & 0.36 &
             $1.6\,10^{-10}$ & $2.3\,10^{-11}$ &
             $1.3\,10^{-1}$ & $\sim 20$ \\
         \hline
      \end{tabular}
      \label{tab:summary}
   \end{minipage}
\end{table*}

\subsection{Disc Evolution under the Influence of the Neutron Star}
\label{sec:sd_binary}

\citet{al94} investigated the tidal/resonant truncation of
circumstellar and circumbinary discs in eccentric binaries and found that
a gap is always formed between the disc and the binary orbit.
Following their formulation, \citet{no01} and \citet{on01a} showed
that the Be disc in Be/X-ray binaries is truncated via the resonant
interaction with the neutron star as long as $\alpha_{\rm SS} \ll 1$.
In the following, we investigate the resonant interaction between the
Be disc and the neutron star in more detail, by analysing the results
from 3D SPH simulations.  Table~\ref{tab:summary} lists some
characteristic quantities from these simulations, which will be
discussed below.

\begin{figure}
  \resizebox{\hsize}{!}{\includegraphics{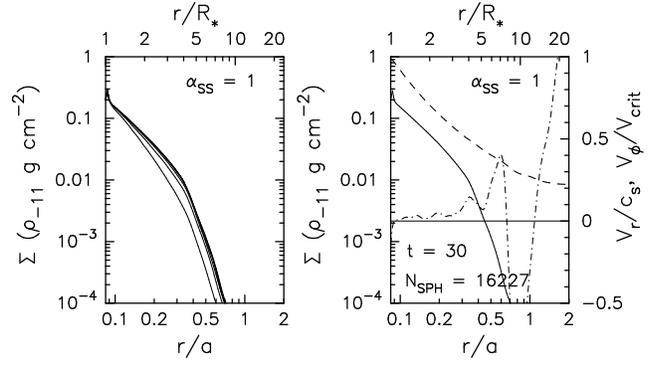}}
  \caption{Evolution of the viscous decretion disc with
    $\alpha_{\rm SS}=1$ in a Be/X-ray with $P_{\rm orb}=24.3\,{\rm d}$
    and $e=0.34$.  The left panel shows the surface density evolution.
    The time interval between adjacent contours is 5$P_{\rm orb}$
    ($\sim 1/3\,{\rm yr}$) ($t =$ 5$P_{\rm orb}$, 10$P_{\rm orb}$,
    $\ldots$ from left).  The right panel shows the disc structure at
    the end of the simulation. The solid, the dashed, and the
    dash-dotted lines denote the surface density, the azimuthal
    velocity normalised by the critical velocity of the Be star, and
    the radial Mach number.  In both panels, the density is integrated
    vertically and averaged azimuthally, while the velocity components
    are averaged vertically and azimuthally. The profile of $V_{\phi}$
    for $r \la 0.7a$ is indistinguishable from one proportional to
    $r^{-1/2}$. Annotated at the bottom of the right panel is
    the number of SPH particles, $N_{\rm SPH}$, at this epoch.}
   \label{fig:sd_a1}
\end{figure}

\begin{figure}
  \resizebox{\hsize}{!}{\includegraphics{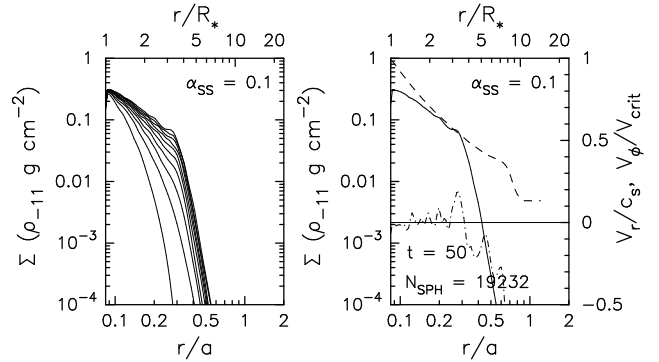}}
  \caption{Same as Fig.\ref{fig:sd_a1}, but for
    $\alpha_{\rm SS}=0.1$.}
   \label{fig:sd_a01}
\end{figure}

Figs.~\ref{fig:sd_a1} and \ref{fig:sd_a01} show the disc evolution
under the influence of the neutron star for $\alpha_{\rm SS}=1$ and
$\alpha_{\rm SS}=0.1$, respectively. The truncation of the disc is
obviously more evident for $\alpha_{\rm SS}=0.1$ than for $\alpha_{\rm
  SS}=1$.  From Fig.~\ref{fig:sd_a01}, we clearly see how the resonant
torque works on a viscous decretion disc.  When the disc size is small
($t < 10P_{\rm orb}$ for $\alpha_{\rm SS}=0.1$), the disc evolution is
almost identical to that around an isolated Be star. As the disc
grows, however, the effect of the resonant torque from the neutron
star becomes apparent and the radial density distribution begins to
have a break at a radius around the 4:1 to 5:1 resonance radii (for
$\alpha_{\rm SS}=0.1$). We call this radius the truncation radius.
Since the resonant torque prevents disc material from drifting
outwards, the disc density increases more rapidly than in discs around
isolated Be stars. Outside the truncation radius, the density
decreases rapidly. The wavy patterns seen in the surface density
distribution in the left panel and in the radial velocity distribution
in the right panel of Fig.~\ref{fig:sd_a01} are due to the
tightly wound spiral density wave excited in the disc.

In contrast to the low-viscosity simulation shown in
Fig.\ref{fig:sd_a01}, the resonant torque has little effect on the
disc structure when the viscosity is very high. As seen in the left
panel of Fig.\ref{fig:sd_a1}, there is an only very modest break in
the surface density distribution for $\alpha_{\rm SS}=1$. In this
simulation, the disc almost reaches an equilibrium state at $t \sim
15P_{\rm orb}$ ($\sim 1\,{\rm yr}$) in the sense that the disc
structure varies regularly, depending on the orbital phase.
  
We also performed a simulation with $\alpha_{\rm SS}=0.3$. The disc
structure obtained is something between those with $\alpha_{\rm
  SS}=0.1$ and $\alpha_{\rm SS}=1$. There is a clear break and a wavy
feature in the radial surface-density distribution but they are not so
strong as those for $\alpha_{\rm SS}=0.1$.

\begin{figure}
  \resizebox{\hsize}{!}{\includegraphics{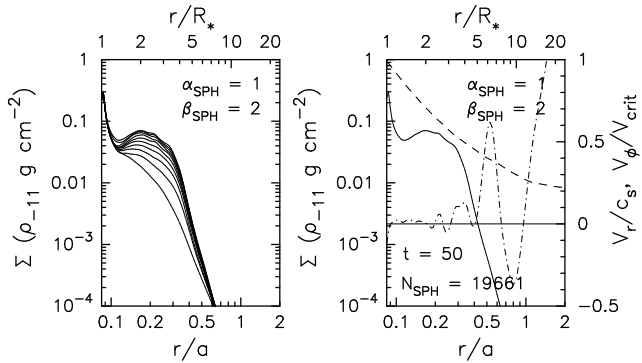}}
  \caption{Same as Fig.~\ref{fig:sd_a1}, but for
    $\alpha_{\rm SPH}=1$ and $\beta_{\rm SPH}=2$.}
   \label{fig:sd_copl2}
\end{figure}

For comparison, we present in Fig.~\ref{fig:sd_copl2} the result from
a simulation with $\alpha_{\rm SPH}=1$ and $\beta_{\rm SPH}=2$, in
which $\alpha_{\rm SS}$ is variable in time and space.  From
Fig.~\ref{fig:sd_copl2}, we observe that the disc with constant
artificial viscosity parameters evolves in a similar way to that of
the $\alpha$-disc with a similar viscosity parameter, except for the
presence of the dip in the density distribution close to the star, as
was expected from the previous section.

\begin{figure}
  \resizebox{\hsize}{!}{\includegraphics{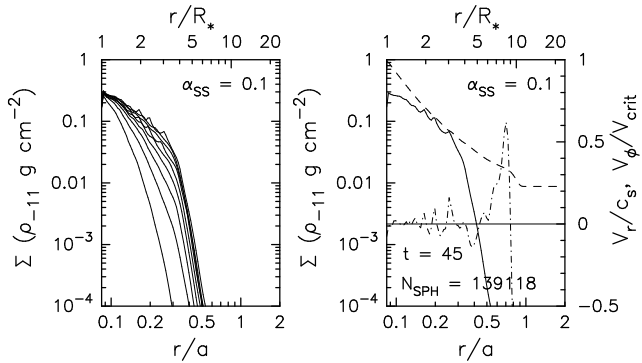}}
  \caption{Same as Fig.~\ref{fig:sd_a01}, but
    for the high-resolution simulation.}
   \label{fig:sd_ukaff}
\end{figure}

In order to study the interaction between the Be disc and the neutron
star in more detail, we performed a high-resolution simulation with
$\alpha_{\rm SS}=0.1$. In this simulation, the number of SPH particles
is about an order of magnitude larger than, and so the smoothing
length is on average less than a half of, that of other simulations.
Fig.~\ref{fig:sd_ukaff} shows the surface density evolution for
$t=0-45P_{\rm orb}$ and the disc structure at $t=45P_{\rm orb}$ in
this high-resolution simulation (unfortunately,
the allocated computing time ran out at $t=47P_{\rm orb}$). From
Fig.~\ref{fig:sd_ukaff}, we easily see more detailed disc structure
than that in the corresponding simulation with a lower resolution
shown in Fig.~\ref{fig:sd_a01}.  The wavy pattern in the surface
density distribution caused by the spiral density wave is more
remarkable in the high-resolution simulation.  This is because a
larger number of particles give a higher resolution of the interacting
region, which makes the interaction more localised and stronger.
Although the surface density profile has breaks near the 5:1 resonance
radius ($r \sim 0.33a$) and the 4:1 radius ($r \sim 0.39a$), the steep
density decrease already begins at a much smaller radius, which
coincides with the outermost density peak of the spiral wave.

In the rest of this paper, we mainly present the results from this
high-resolution simulation as the representative ones with $\alpha_{\rm
  SS}=0.1$, because it gives a better understanding of the star-disc
interaction.

\subsection{Phase-Dependent Disc Structure}
\label{sec:phase}

Most Be/X-ray binaries with known orbital parameters have orbital
eccentricities in the range from 0.3 to 0.9. In such systems, the
star-disc interaction is likely to be strongly phase-dependent. In
this subsection, we discuss phase-dependent features except for the
mass capture rate by the neutron star, which will be discussed
separately.

\begin{figure*}
  \resizebox{\hsize}{!}
  {\includegraphics[angle=-90]{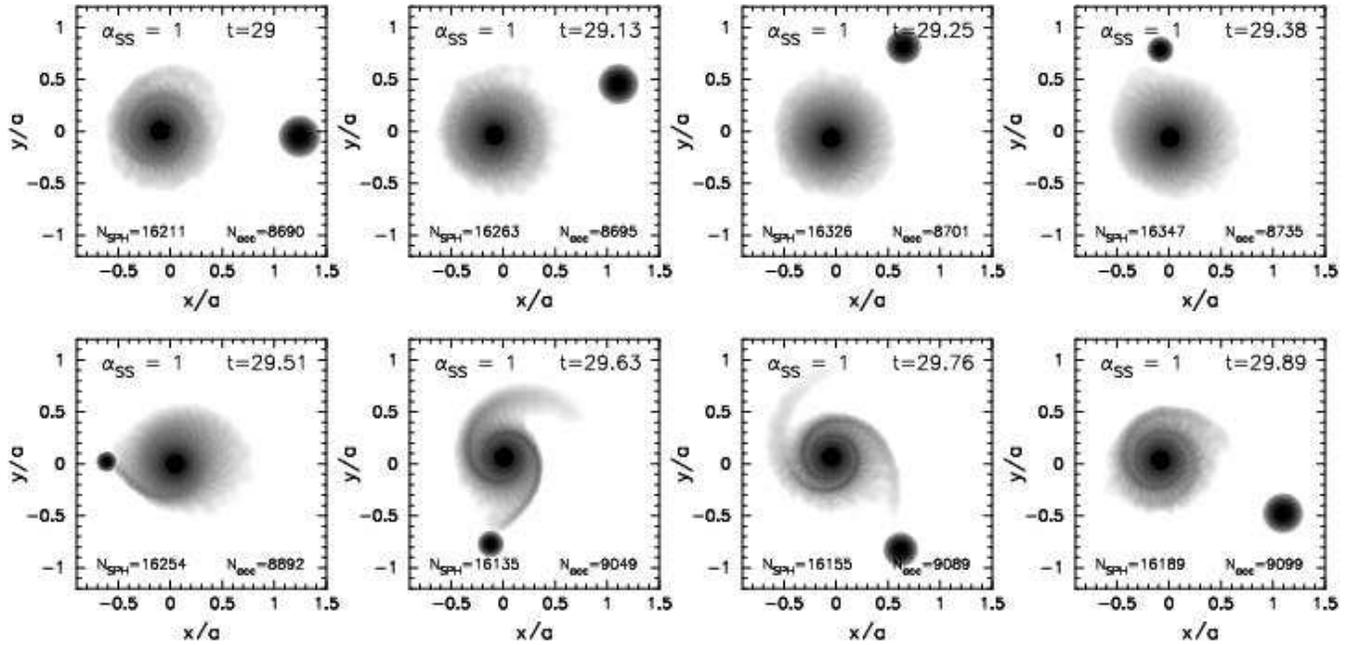}}
  \caption{Snapshots for $\alpha_{\rm SS}=1$, which cover one 
    orbital period.  Each panel shows the logarithm of the surface
    density. The dark spot near the origin is the Be star, while
    another dark spot orbiting about the Be star denotes the neutron
    star with the variable accretion radius. Annotated at the
      bottom of each panel are the number of SPH particles, $N_{\rm
        SPH}$, and the integrated number of particles captured by the
      neutron star, $N_{\rm acc}$.}
   \label{fig:ss_a1}
\end{figure*}

\begin{figure*}
  \resizebox{\hsize}{!}
  {\includegraphics[angle=-90]{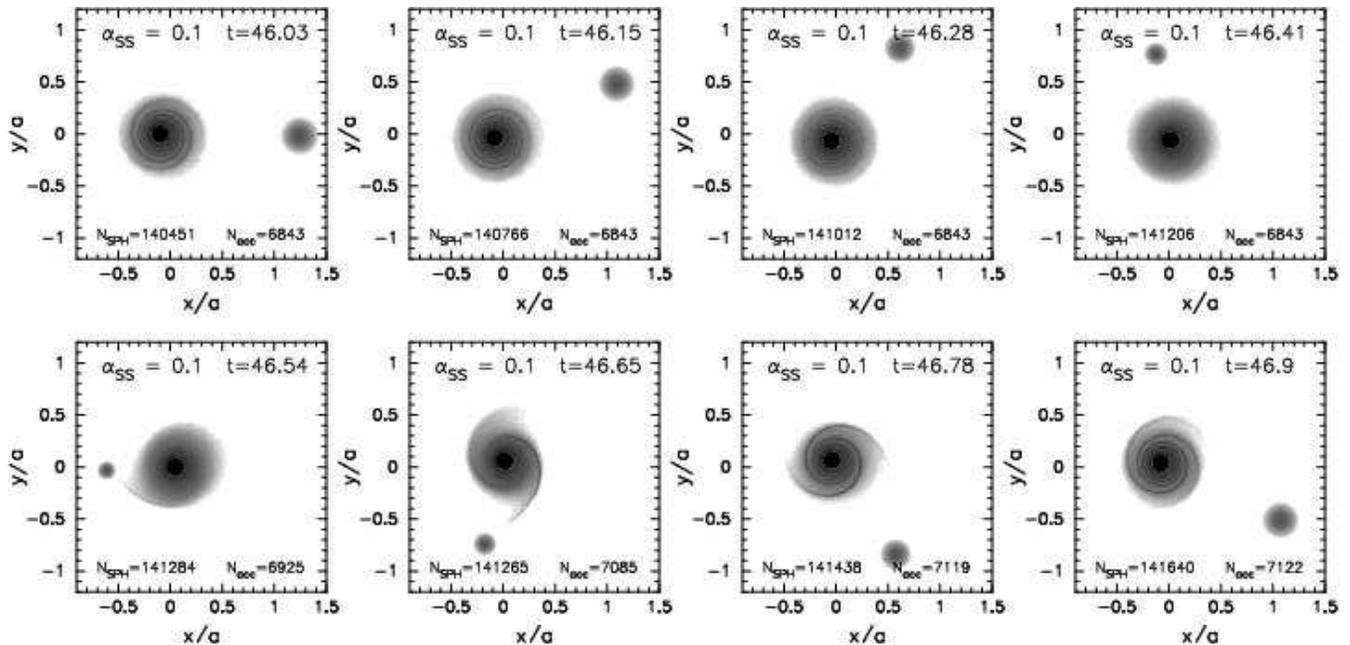}}
  \caption{Same as Fig.\ref{fig:ss_a1}, but for 
    the high-resolution simulation with $\alpha_{\rm SS}=0.1$.}
   \label{fig:ss_a01}
\end{figure*}

Figs.\ref{fig:ss_a1} and \ref{fig:ss_a01} give snapshots covering one
orbital period for $\alpha_{\rm SS}=1$ and $\alpha_{\rm SS}=0.1$,
respectively. Each panel shows the surface density in a range of about
two orders of magnitude in the logarithmic scale.  From these figures,
we note a remarkable difference in the disc structure between
$\alpha_{\rm SS}=1$ and $\alpha_{\rm SS}=0.1$. For $\alpha_{\rm
  SS}=1$, the disc has a significant density up to the periastron
distance and experiences a strong interaction at and after the
periastron passage of the neutron star.

On the other hand, for $\alpha_{\rm SS}=0.1$, the resonant torque from
the neutron star is much more effective at truncating the disc than
for the high viscosity disc.  The sharp decline in the disc density
outside the 5:1 resonance causes a gap between this radius and the
periastron distance, apparently reducing the mass capture rate by the
neutron star, as will be seen in Section~\ref{sec:capture}.

In both cases, the spiral density waves are clearly seen between the
periastron passage and the apastron passage. The opening angle of 
the spirals, which is related to the effective gravity in the disc,
is smaller for a larger binary separation.

Through the resonant interaction, the angular momentum is
transported from the disc to the binary. We have to admit that
the effect of the angular momentum transport on the binary turned
out to be much stronger than we had expected. Despite the fact that
the mass of each particle is only $10^{-10}M_{\sun}$ so that the disc
mass is only about $10^{-5}-10^{-6}M_{\sun}$ in our simulations, the
increase in the binary orbital period is visible in the late stage
of simulations. The computed phase lags behind the correct orbital
phase by $\sim 0.01$ at $t \sim 30$ in the simulation with
$\alpha_{\rm SS}=1$ and by $\sim 0.04$ and $\sim 0.05$ at $t \sim
45$ in the high and normal resolution simulations with $\alpha_{\rm
  SS}=0.1$, respectively. In the following, figures should be read
taking this phase shift into account.

In order to have a measure of the disc radius, i.e., the radius at
which the disc density has a major break, we applied a non-linear
least-square fitting method to the radial distribution of the
azimuthally-averaged surface density $\Sigma$, adopting the following
simple fitting function,
\begin{equation}
 \Sigma \propto \frac{\left( r/r_{\rm d}
   \right)^{-p}}{1+\left( r/r_{\rm d} \right)^{q}},
   \label{eq:lsq}
\end{equation}
where $p$ and $q$ are constants and $r_{\rm d}$ is the disc radius.
 
\begin{figure}
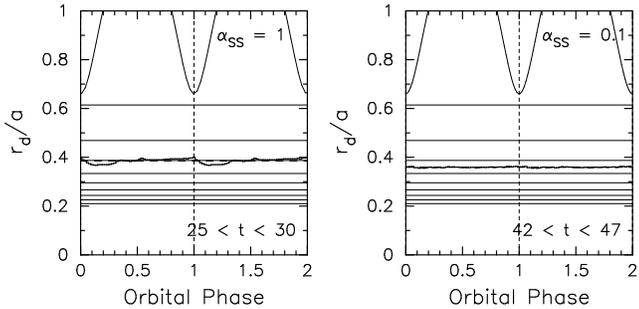

  \resizebox{\hsize}{!}{\includegraphics{fig12a.eps}
    \quad\quad \includegraphics{fig12b.eps}}
    \caption{Orbital-phase dependence of the disc radius $r_{\rm d}$.
      Averaging is done over $25 \le t \le 30$ for $\alpha_{\rm SS}=1$
      (left) and $42 \le t \le 47$ for $\alpha_{\rm SS}=0.1$
      (right). The horizontal solid lines denote the 2:1, the 3:1, the
      4:1, $\ldots$, the 10:1 resonance radii from top to bottom, and
      the horizontal dashed line denotes the phase averaged value of
      $r_{\rm d}$. The thin sinusoidal line denotes the orbit of the
      neutron star. The periastron passage of the neutron star, which
      occurs at phase 0, is denoted by the vertical dashed line.}
    \label{fig:rd_copl}
\end{figure}

Fig.~\ref{fig:rd_copl} shows the phase dependence of the disc radius
(thick line) in the simulations shown in Figs.\ref{fig:ss_a1} and
\ref{fig:ss_a01}. To reduce the fluctuation noise, we folded the
data on the orbital period over $25 \le t \le 30$ for $\alpha_{\rm
  SS}=1$ and $42 \le t \le 47$ for $\alpha_{\rm SS}=0.1$.  In the
figure, the horizontal solid lines denote some of the lowest $n:1$
resonance radii (the 2:1, the 3:1, $\ldots$, the 10:1 from top), while
the thin sinusoidal line denotes the orbit of the neutron star. The
origin of the phase is at the periastron passage of the neutron star.

From Fig.~\ref{fig:rd_copl}, we note that the disc radius coincides
with the 4:1 resonance radius ($r/a = 0.39$) for $\alpha_{\rm SS}=1$,
whereas the disc has a radius intermediate between the 4:1 radius and
the 5:1 radius ($r/a = 0.33$) for $\alpha_{\rm SS}=0.1$ (the mean
$r_{\rm d}$ is $0.39a$ for $\alpha_{\rm SS}=1$ and $0.36a$ for
$\alpha_{\rm SS}=0.1$). The latter is a typical feature expected for
a disc in which the wave damps locally \citep{al94}.

We also note that the disc radius modulates around the mean value.
The disc shrinks after the periastron passage of the neutron star,
which gives a negative torque to the disc. After that it restores its
radius by viscous diffusion, so that the amplitude of the modulation
is larger for $\alpha_{\rm SS}=1$ than for $\alpha_{\rm SS}=0.1$.

Since the disc structure depends on the binary phase, the disc
emission is expected to exhibit an orbital modulation as well. To see
whether this is the case, we calculated the continuum flux from the Be
disc for $r \ge 2r_{\rm inj}-R_{*}$, assuming that the disc is
optically thin and the emissivity is proportional to $\rho^2$, where
$\rho$ is the local density. For simplicity, we ignored the effect of
the obscuration of the disc by the star, which will become important
for systems with high inclination angles. We then obtained a base flux
curve by performing the cubic-spline fitting of the fluxes at apastron
passages. The base flux describes the long-term change in the
continuum flux. Finally, we computed the orbital modulation by
subtracting the base flux from the instantaneous fluxes.

\begin{figure}
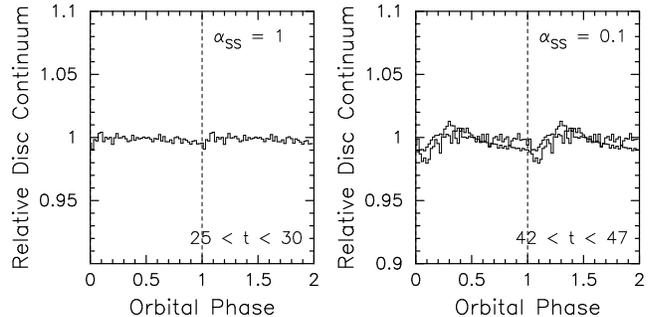

  \resizebox{\hsize}{!}{\includegraphics{fig13a.eps}
    \quad\quad \includegraphics{fig13b.eps}}
    \caption{Orbital modulation in the optically thin continuum
      from the Be disc for $\alpha_{\rm SS}=1$ (left) and
      $\alpha_{\rm SS}=0.1$ (right). In the right panel, the thick and
        thin lines are for the high and normal resolution simulations,
        respectively. Phase 0 corresponds to the periastron passage
      of the neutron star.}
    \label{fig:lc_copl}
\end{figure}

Fig.~\ref{fig:lc_copl} shows the orbital modulation in the optically
thin continuum from the Be disc for $\alpha_{\rm SS}=1$ (left) and
$\alpha_{\rm SS}=0.1$ (right).  For $\alpha_{\rm SS}=0.1$, the
modulation is more clearly seen in the high-resolution simulation
(thick line) than in the normal one (thin line).  In order to reduce
the fluctuation noise, we folded the data on the orbital period
over the period annotated in the panel.  Contrary to what is expected
from Fig.\ref{fig:rd_copl}, the continuum exhibits little modulation
for $\alpha_{\rm SS}=1$, and about one per cent of modulation is seen
for $\alpha_{\rm SS}=0.1$. The negative result for $\alpha_{\rm SS}=1$
suggests that the strongly-perturbed outer disc contributes little to
the optically thin emission, because of its low density. On the other
hand, the positive result for $\alpha_{\rm SS}=0.1$, in particular in
the high-resolution simulation, though the amplitude is still very
small, seems to come from the region where the density is enhanced by
the spiral wave, because the disc radius does not change significantly
and the rise and the subsequent decline of disc emission are in phase
with the density enhancement and the subsequent damping caused by the
spiral density wave.  The difference between the modulation patterns
from the high and normal resolution simulations with $\alpha_{\rm
  SS}=0.1$ suggests that resolving the spiral density wave is
important to obtain a reliable modulation pattern of such low amplitude.

Although Fig.~\ref{fig:lc_copl} gives little observability of the
orbital modulation in optically thin disc emission, it is still likely
that optically thick emission, such as Balmer lines, will show
significant orbital modulation. Calculating the optically thick
emission from the disc is, however, beyond the scope of this paper.

\subsection{Excitation of the Eccentric Mode}
\label{sec:ecc_mode}

In a circumbinary disc around an eccentric binary, an eccentric mode
is excited through direct driving due to a one-armed bar potential
\citep{al96a}.  In this subsection, we study whether the same mechanism
works in Be/X-ray binaries. We analyse the evolution of the
eccentricity in the Be disc by decomposing the surface density
distribution into Fourier components which vary as $\exp i(k \phi -
\ell \Omega_{\rm orb} t)$, where $k$ and $\ell$ are the azimuthal and
time-harmonic numbers, respectively, and $\Omega_{\rm orb} =
[G(M_{*}+M_{X})/a^3]^{1/2}$ is the frequency of mean binary rotation.
 
Following \citet{lub91}, we define the mode strength by
\begin{eqnarray}
   S_{f, g, k, l} &=& \frac{2}{\pi M_{\rm d} (1+\delta_{k, 0})
         (1+\delta_{\ell, 0})}
         \int_t^{t+2\pi \Omega_{\rm orb}^{-1}} dt'
         \nonumber\\
         &&\nonumber\\
         && \times \int dr \int_0^{2\pi} d\phi\ 
            \Sigma(r, \phi, t) f(k \phi) g(\ell t'),
   \label{eq:sfgkl}
\end{eqnarray}
where $f$ and $g$ are either $\sin$ or $\cos$ functions.  The surface
density here is computed by summing up $\delta$ functions at particle
positions, not by taking the kernel into account as has been done in
the previous sections.  Then, the total strength of the mode $(k,
\ell)$ is defined by
\begin{eqnarray}
   S_{k, \ell}(t) &=& (S_{\cos, \cos, k, \ell}^2
          + S_{\cos, \sin, k, \ell}^2
         \nonumber\\
         && + S_{\sin, \cos, k, \ell}^2
          + S_{\sin, \sin, k, \ell}^2)^{1/2}.
   \label{eq:skl}
\end{eqnarray}
 
\begin{figure}
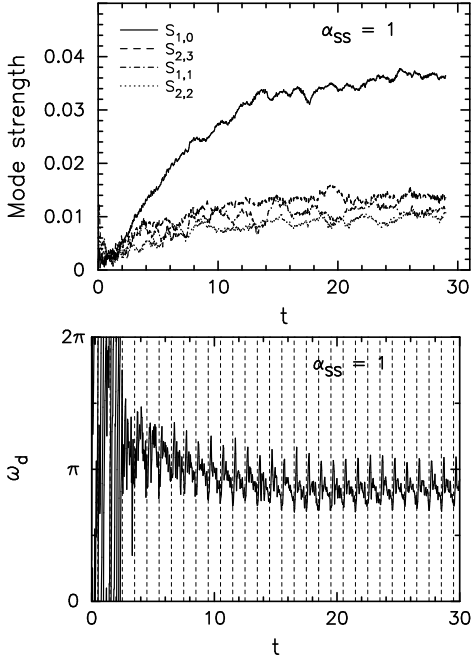

  \centerline{\includegraphics[width=0.35\textwidth]
    {fig14a.eps}}
  \centerline{\includegraphics[width=0.35\textwidth]
    {fig14b.eps}}
    \caption{Excitation of the eccentric mode. The upper panel
      shows the strengths of several modes.  The solid, the dashed,
      the dash-dotted, and the dotted lines denote the strengths of
      the (1,0) mode, the (2,3) mode, the (1,1) mode, and the (2,2)
      mode, respectively. The lower panel shows the angle between the
      eccentric vectors of the disc and the binary.}
   \label{fig:mode_a1}
\end{figure}

Fig.~\ref{fig:mode_a1} shows the excitation and precession of the
(1,0) mode (i.e., the eccentric mode) for $\alpha_{\rm SS}=1$. The
upper panel shows the strengths of several modes, while the lower
panel shows the evolution of the angle $\omega_{\rm d}$ between the
eccentric vector of the disc and that of the binary orbit defined by
\begin{equation}
   \tan \omega_{\rm d} = \frac{\int dr \int_0^{2\pi} d\phi\
            \Sigma(r, \phi, t) \sin \phi}
            {\int dr \int_0^{2\pi} d\phi\
            \Sigma(r, \phi, t) \cos \phi}.
   \label{eq:omega_d}
\end{equation}

The eccentric mode grows initially linearly in time ($t \la 8$), as is
predicted by the theory and was also found by \citet{la00}.
At $t \sim 15$ the strength of the eccentric mode is saturated at
$S_{1,0} \sim 0.04$. As it is saturated, the mode stops precessing and
is locked at $\omega_{\rm d} \sim \pi$.

In order to study the structure of the eccentric mode in more detail,
we analyse the orbits of individual particles. The position and
velocity of each particle are instantaneously equal to those of an
elliptical Keplerian orbit of semi-latus rectum $\lambda$ and the
eccentric vector $\bmath{e}_{\rm SPH}$ with the amplitude $e_{\rm
  SPH}$ and the longitude of periastron, $\omega$, with respect to the
stellar periastron (e.g., \citealt{ogi01}). The radial coordinate $r$
is then given by
\begin{equation}
   r = \frac{\lambda}{1+e_{\rm SPH} \cos (\phi-\omega)}.
\end{equation}
Given $e_{\rm SPH} \ll 1$, $\lambda \sim r$ in our simulations.

\begin{figure}
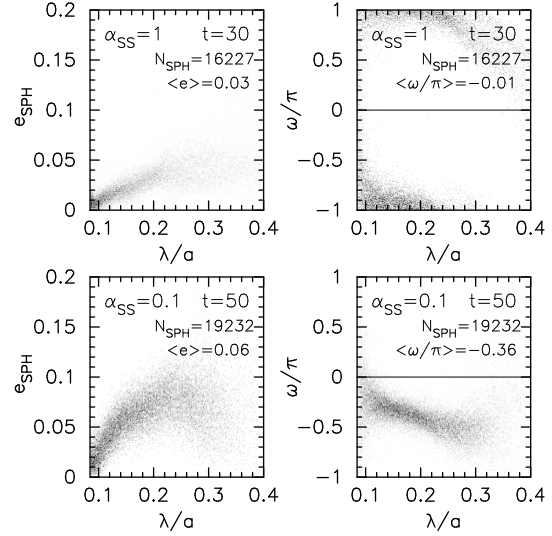

  \centerline{\includegraphics[width=0.4\textwidth]
    {fig15a.eps}}
  \centerline{\includegraphics[width=0.4\textwidth]
    {fig15b.eps}}
   \caption{Eccentricity $e_{\rm SPH}$ and the longitude of
     periastron, $\omega$, of the particle orbits from the
     normal-resolution simulations with $\alpha_{\rm SS}=1$ at $t=30$
     (upper) and $\alpha_{\rm SS}=0.1$ at $t=50$ (lower).  The
     gray-scale plot in each panel shows the particle distribution in
     linear scale. $<e>$ and $<\omega/\pi>$ are mean values of $e_{\rm
     SPH}$ and $\omega/\pi$.}
   \label{fig:ecc_lowres}
\end{figure}

The upper panels of Fig.~\ref{fig:ecc_lowres} present the distribution
of the orbital parameters of disc particles for $\alpha_{\rm SS}=1$ at
$t=30$. From the left panel, we note that the disc eccentricity
increases almost linearly in $\lambda$ for $\lambda \la 0.3a$ and is
roughly constant for $\lambda \ga 0.3a$. The right panel for the
radial dependence of the longitude of periastron shows that the
eccentric mode is twisted in a trailing sense.

The normal-resolution simulation with $\alpha_{\rm SS}=0.1$ showed a
similar trend. The eccentric mode grew initially linearly in time ($t
\la 40$). The mode strength was almost saturated at $t \sim 45$ at
$S_{1,0} \sim 0.1$.  The mode also stopped precessing and was locked
at $\omega_{\rm d} \sim 3\pi/2$.
  
The distribution of the orbital parameters of disc particles at $t=50$
from this simulation is presented in the lower panels of
Fig.~\ref{fig:ecc_lowres}. From the left panel, we note that the
eccentricity has a maximum at $\lambda \sim 0.3a$. The right panel
shows that the eccentric mode is twisted in a trailing sense, as in
the case for $\alpha_{\rm SS}=1$.

Recently, \citet{ogi01} has developed a non-linear theory of the
evolution of the shape and surface density of a three-dimensional
eccentric accretion disc.  When the eccentricity of the disc is small
so that terms of relative order $O(e^2)$ may be neglected, this theory
provides a linear evolutionary equation for the complex eccentricity
$E(\lambda,t)=e(\lambda,t)\,{\rm e}^{{\rm i}\omega(\lambda,t)}$ of the
disc.  We have reworked this theory for the case of a strictly
isothermal decretion disc with no radial velocity, and also evaluated
the tidal forcing terms associated with a companion object of small
eccentricity.  The governing equation is then of the form
\begin{eqnarray}
  \lefteqn{\Sigma(GM_*\lambda)^{1/2}{{\partial E}\over{\partial t}}=
  {{\partial}\over{\partial\lambda}}(Z_1\Sigma c_{\rm s}^2\lambda)+
  Z_2\Sigma c_{\rm s}^2}&\nonumber\\
  &&+{{1}\over{4}}\frac{{\rm i}GM_X\Sigma\beta}{\lambda_{\rm b}}
  \left[b_{3/2}^{(1)}(\beta)E-b_{3/2}^{(2)}(\beta)E_{\rm b}\right],
  \label{eq:dedt}
\end{eqnarray}
where $E_{\rm b}$ and $\lambda_{\rm b}$ are the (constant)
eccentricity and semi-latus rectum of the binary orbit,
$b_\gamma^{(m)}$ is the Laplace coefficient of celestial mechanics,
and $\beta=\lambda/\lambda_{\rm b}$.  The dimensionless coefficients
$Z_1$ and $Z_2$ are given by
\begin{equation}
  Z_1=C_1E+C_2\lambda{{\partial E}\over{\partial\lambda}},\qquad
  Z_2=C_3E+C_4\lambda{{\partial E}\over{\partial\lambda}},
\end{equation}
where $C_1,\dots,C_4$ are dimensionless complex constants that depend
only on the shear and bulk viscosity parameters $\alpha_{\rm SS}$ and
$\alpha_{\rm b}$. In the limit $\alpha_{\rm SS}, \alpha_{\rm b} \to 0$
these coefficients become purely imaginary and give rise to precession
induced by the pressure of the disc. Most importantly, the real part
of $C_2$, which has the role of a diffusion coefficient for
short-wavelength eccentric perturbations, turns out to be positive
when $\alpha_{\rm b}/\alpha_{\rm SS}>2/3$, as is true in the SPH
simulations [see equation~(\ref{eq:nubsph})].  If this condition were not
satisfied, the disc would experience a local eccentric instability and
equation (\ref{eq:dedt}) could not be evolved forward in time.

Equation (\ref{eq:dedt}) has the character of a dispersive and
diffusive linear wave equation for $E(\lambda,t)$.  The eccentricity
of the binary provides a tidal forcing that is independent of time.
Starting from an initially circular state $E=0$, the eccentricity of
the disc first grows linearly in time and then approaches a steady
state.  The steady eccentric shape can be determined either by solving
the time-dependent problem until the transient response decays, or by
solving the second-order ordinary differential equation obtained by
setting the time-derivative to zero.  We have performed both of these
calculations, adopting the boundary condition $E=0$ at the stellar
surface, a `stress-free' boundary condition $Z_1=0$ at a notional
outer edge ($\beta=0.39$), and a surface density
distribution $\Sigma\propto\lambda^{-2}$.

\begin{figure}
  \centerline{\includegraphics[width=0.4\textwidth]
    {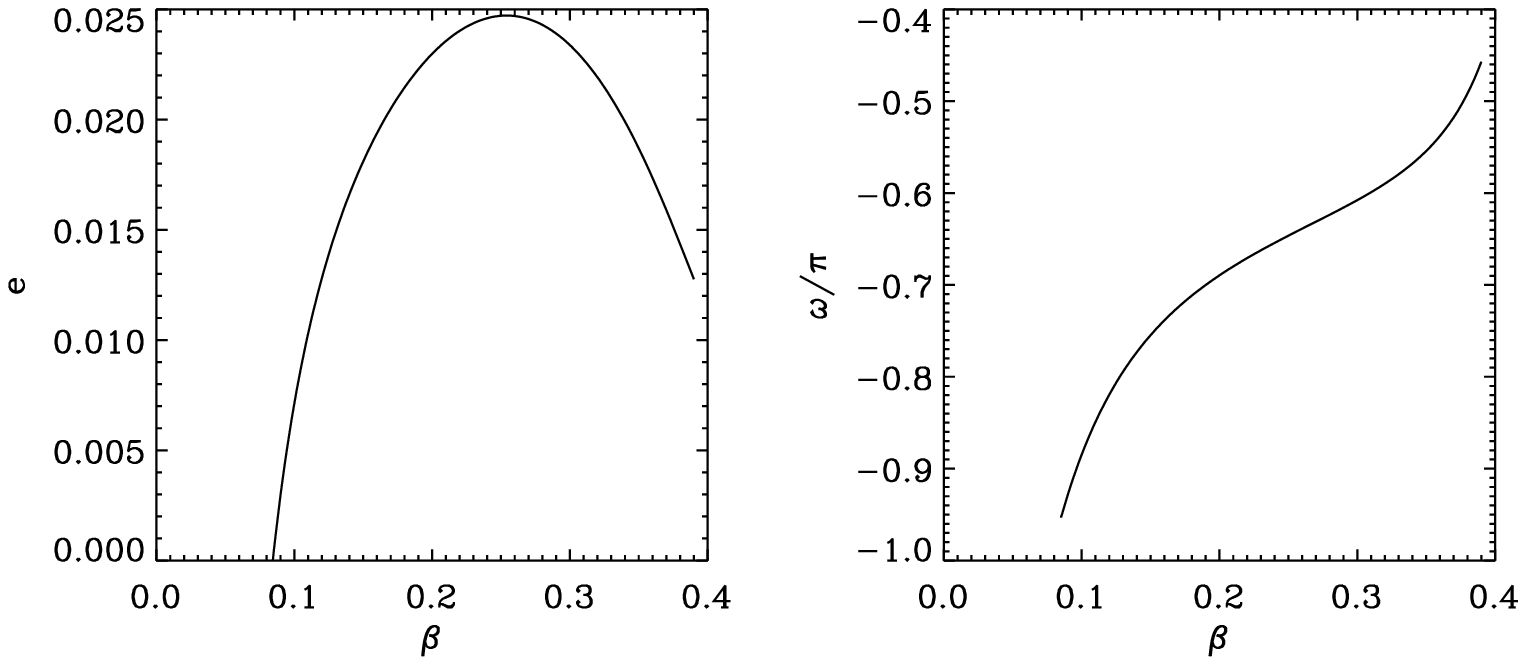}}
  \centerline{\includegraphics[width=0.4\textwidth]
    {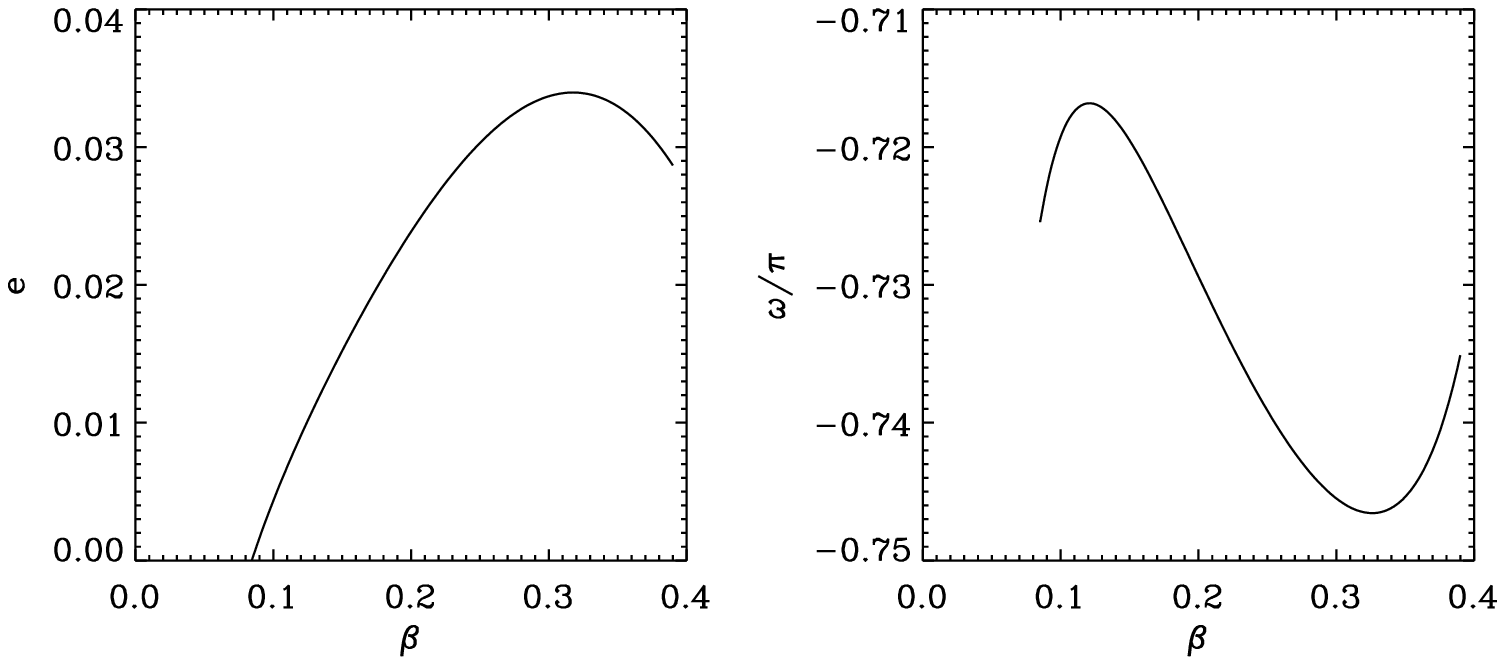}}
   \caption{Model eccentricity $e$ and longitude of periastron
     $\omega$ of the particle orbits for $\alpha_{\rm SS}=1$ (upper)
     and $\alpha_{\rm SS}=0.1$ (lower).  The notional outer edge is at
     $\beta (=\lambda/\lambda_{\rm b})=0.39$.}
   \label{fig:ecc_model}
\end{figure}

The steady configurations of $e(\lambda)$ and $\omega(\lambda)$ for
the cases $\alpha_{\rm SS}=1$ and $\alpha_{\rm SS}=0.1$ are
illustrated in Fig.~\ref{fig:ecc_model}.  One should not expect a
perfect agreement with the SPH simulations because the theory is based
on the assumption of small eccentricities and only approximates the
actual surface density distribution.  However, the steady eccentric
shapes based on linear theory offer a fair explanation of the results
of the SPH simulations, especially for the high-viscosity case.

\begin{figure}
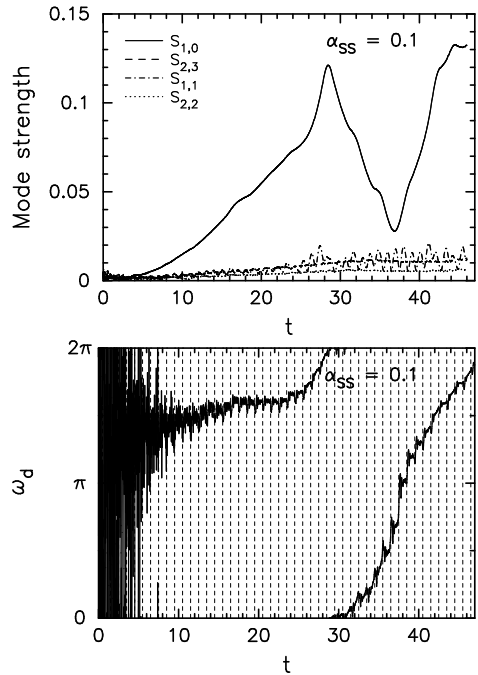

  \centerline{\includegraphics[width=0.35\textwidth]
    {fig17a.eps}}
  \centerline{\includegraphics[width=0.35\textwidth]
    {fig17b.eps}}
   \caption{Same as Fig.~\ref{fig:mode_a1}, but for
     the high-resolution simulation with $\alpha_{\rm SS}=0.1$.}
   \label{fig:mode_a01}
\end{figure}

\begin{figure*}
  \resizebox{\hsize}{!}{\includegraphics{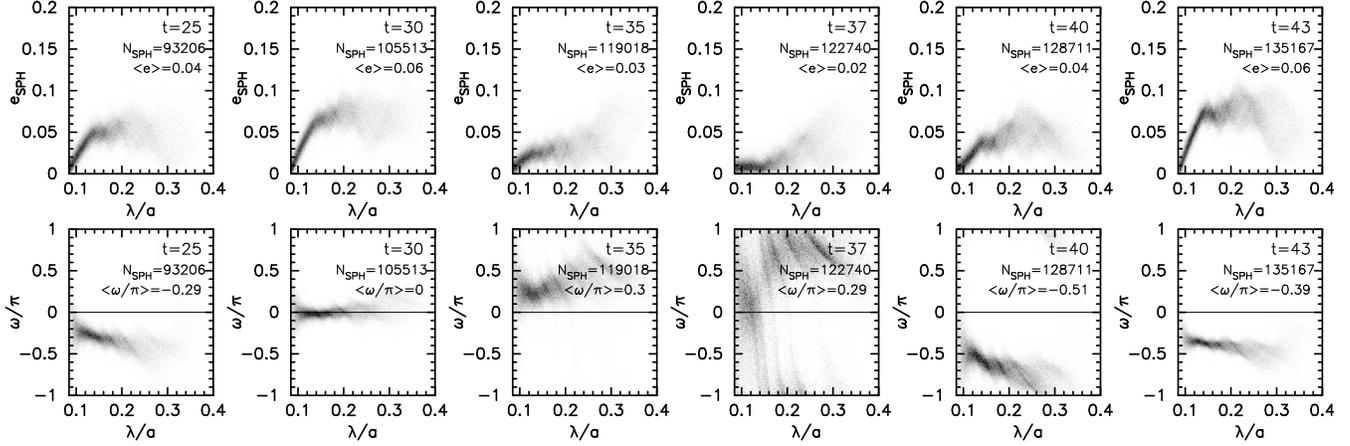}}
   \caption{Change in the eccentricity $e_{\rm SPH}$ and the longitude
     $\omega$ of the disc particle orbits in the high-resolution
     simulation with $\alpha_{\rm SS}=0.1$. Epochs are chosen for
     illustrative purpose, so the time interval is not
     constant. The gray-scale plot in each
     panel shows the particle distribution in linear scale. $<e>$ and
     $<\omega/\pi>$ are mean values of $e_{\rm SPH}$ and $\omega/\pi$
     at each epoch, respectively.}
   \label{fig:ecc_a01hi}
\end{figure*}

Fig.~\ref{fig:mode_a01} shows the evolution of an eccentric mode from
the high-resolution simulation with $\alpha_{\rm SS}=0.1$.  The
detailed structure of the mode is shown in Fig.~\ref{fig:ecc_a01hi} by
snapshots covering one precessional period.  From these figures, we
note that the evolution and the structure of the eccentric mode are
very similar to those from the corresponding normal resolution
simulation until $t \sim 25$. At $t \sim 25$, the eccentric mode
suddenly begins to precess in a prograde sense, as seen in the lower
panel of Fig.\ref{fig:mode_a01}. As $\omega_{\rm d}$ increases from
$\sim 3\pi/2$ to $2\pi$, the mode strength increases.  The mode
stagnates at $\omega_{\rm d} \sim 0$ for $(1-2)P_{\rm orb}$ around $t
\sim 30$, suggesting that the eccentric disc at $\omega_{\rm d} \sim
0$ is an unstable configuration. The precession decelerates before
reaching $\omega_{\rm d} = 0$ and accelerates after that. At the same
time, the mode changes its shape. As seen in the lower panel of
Fig.~\ref{fig:ecc_a01hi}, the eccentric mode is twisted in a trailing
sense for $\omega_{\rm d} < 0$ and in a leading sense after that. At
$\omega_{\rm d}=0$, apsidal axes of all particle orbits align.
  
As seen in Fig.~\ref{fig:ecc_a01hi}, the twist of the mode increases,
as the apsidal axis of the disc moves toward the stellar periastron
($\omega_{\rm d}=0 \to \pi$). The strong radial dependence of the
phase caused by the twist reduces the eccentricity of the disc, as
shown in the upper panel of Fig.~\ref{fig:mode_a01}.  The precession
is fastest at $\omega_{\rm d} \sim \pi$, suggesting that there is a
stable point at $\omega_{\rm d} \sim \pi$.  After that, the mode
becomes trailing and its strength increases. The precessional period
is about $20 P_{\rm orb}$.

It is important to note that a similar behaviour is found in
circumbinary discs around eccentric binaries.  According to
\citet{la00}, this behaviour occurs as follows: When the eccentricity
of the disc edge is small, $\omega_{\rm d}$ is locked at a stable
value $\omega_{\rm d} = 3\pi/2$. However, as the eccentricity grows,
the locking action weakens, and the prograde precession due to the
quadrupole moment of the binary potential dominates. The disc edge
begins to precess when its eccentricity becomes $(0.2-0.7) e$, and
afterwards the eccentricity oscillates with a precessional period. The
disc typically attains the eccentricity of $(0.5-1) e$.

We note that the above behaviour of the eccentric mode in circumbinary
discs around eccentric binaries is strikingly similar to that shown in
Fig.~\ref{fig:mode_a01}, except that, in our simulation for Be/X-ray
binaries, the growth and precessional time-scales of the eccentric
mode are much shorter and the disc eccentricity attained is
significantly smaller than in circumbinary discs.

\subsection{Mass Capture Rate by the Neutron Star}
\label{sec:capture}

As mentioned in Section~\ref{sec:intro}, most of the Be/X-ray binaries
show transient X-ray activities. Among them, some exhibit periodical
X-ray outbursts called Type~I, which coincide with the periastron
passage, while the others show occasional giant X-ray outbursts called
Type~II and little or no detectable X-ray emission in quiescent
phase. 4U\,0115+63, the system we are modelling in this paper, belongs
to the latter group.  In this subsection, we first study how much mass
is captured by the neutron star in a general context, and then discuss
whether our model predicts the mass capture rate consistent with the
observed X-ray behaviour of 4U\,0115+63.

\begin{figure}
  \centerline{\includegraphics[width=0.35\textwidth]
    {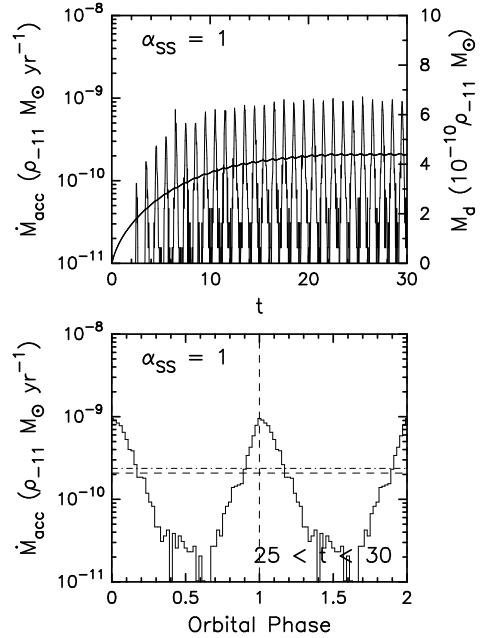}}
    \caption{Evolution of the disc mass and 
      the mass capture rate by the neutron star (upper) and the
      orbital-phase dependence of the mass capture rate (lower). In
      the upper panel, the thin line denotes the mass capture rate
      $\dot{M}_{\rm acc}$ and the thick line denotes the disc mass
      $M_{\rm d}$.  In the lower panel, the data are folded on the
      orbital period over $25 \le t \le 30$.  The horizontal dashed
      line and the dash-dotted line in the lower panel denote the mass
      capture rate by the neutron star and the mass-loss rate from the
      Be star averaged over the period annotated at the bottom of the
      panel.}
    \label{fig:mdot_a1}
\end{figure}

Fig.~\ref{fig:mdot_a1} shows the change in the mass capture rate by
the neutron star, $\dot{M}_{\rm acc}$, and the disc mass $M_{\rm d}$
for $\alpha_{\rm SS}=1$. The upper panel shows the evolution of
$\dot{M}_{\rm acc}$ and $M_{\rm d}$, while the lower panel shows the
orbital-phase dependence of $\dot{M}_{\rm acc}$. In the lower panel,
we folded $\dot{M}_{\rm acc}$ on the orbital period over $25 \le t \le
30$ to reduce the fluctuation noise. The horizontal dashed line and
dash-dotted line denote the mean mass capture rate by the neutron star
and the mean mass-loss rate from the Be star, respectively.

We have already seen that there is little truncation of the Be disc
for $\alpha_{\rm SS}=1$. Fig.~\ref{fig:mdot_a1} confirms this.  For $t
\ga 20$ the Be disc is almost in equilibrium in the sense that the
disc mass and the mass capture rate only shows regular orbital
modulations with constant amplitude.  For $25 \le t \le 30$, the mean
mass-loss rate from the Be star is $2.4 \times 10^{-10} \rho_{-11}
M_\odot{\rm yr}^{-1}$, while the mean mass capture rate for the same
period is $2.1\times 10^{-10} \rho_{-11} M_\odot{\rm yr}^{-1}$.  Thus,
in this high viscosity disc, the neutron star captures the disc mass
at about the same rate as the mass-loss rate from the Be star.

The simulation with $\alpha_{\rm SS}=0.3$ also gives a high fraction
(about 60 per cent for $45 \le t \le 50$) of mass capture rate with
respect to the mass-loss rate, as listed in Table~\ref{tab:summary}.
This indicates that the resonant truncation effect is not effective
for $\alpha_{\rm SS}=0.3$, either.

\begin{figure}
  \centerline{\includegraphics[width=0.35\textwidth]
    {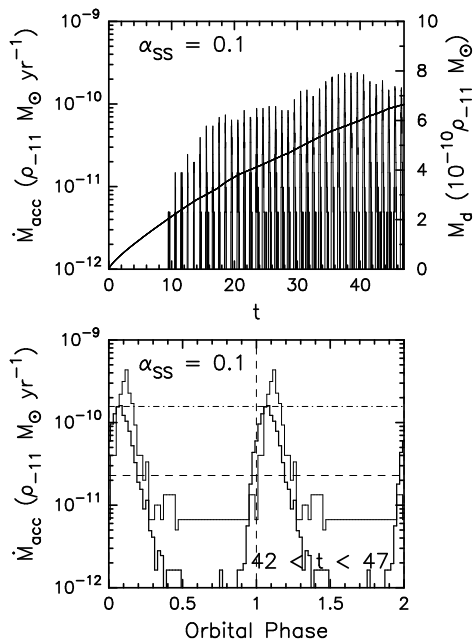}}
    \caption{Same as Fig.~\ref{fig:mdot_a1}, but for
      the high-resolution simulation with $\alpha_{\rm SS}=0.1$. In the
      lower panel, the data are folded on the orbital period over $42
      \le t \le 47$. For comparison, the mass capture rate from the
      normal-resolution simulation with $\alpha_{\rm SS}=0.1$ (thin
      solid line) is overlapped in the lower panel.}
    \label{fig:mdot_a01}
\end{figure}

In contrast to the simulations with $\alpha_{\rm SS}=1$ and
$\alpha_{\rm SS}=0.3$, the high-resolution simulation with $\alpha_{\rm
  SS}=0.1$ revealed a subtle feature in the mass capture rate, as
shown in Fig.~\ref{fig:mdot_a01}.  Before the precession of the
eccentric mode began, the mass capture rate $\dot{M}_{\rm acc}$
increased monotonically: $\dot{M}_{\rm acc} = 0$ before $t \sim 10$
with the resolution of this simulation. Then, the peak mass capture
rate showed a gradual increase to a level at $\dot{M}_{\rm acc} =
(8-9) \times 10^{-11} \rho_{-11} M_{\sun}\,{\rm yr}^{-1}$ at $t \sim
25$.

After the eccentric mode began to precess, $\dot{M}_{\rm acc}$
exhibited a long-term modulation, depending on the longitude of the
eccentric mode, $\omega_{\rm d}$. It gradually increased as
$\omega_{\rm d}$ increased from 0 to $\pi$ and decreased as
$\omega_{\rm d}$ increased from $\pi$ to $2\pi$. At the periastron
passage at $t \sim 38$, $\dot{M}_{\rm acc}$ was a maximum of $(2-3)
\times 10^{-10} \rho_{-11} M_{\sun}\,{\rm yr}^{-1}$, which was about a
factor of three higher than the level before precession. The
eccentricity of the disc was $\sim 0.1$ at $\omega_{\rm d} \sim 0$ and
$0.02-0.03$ at $\omega_{\rm d} \sim \pi$. It should be noted that even
this small eccentricity caused a factor of three enhancement in the
mass capture rate. If the disc had a much stronger disturbance, the
enhancement in $\dot{M}_{\rm acc}$ is likely to be much larger.

In addition to the long-term modulation due to the precession of the
eccentric mode, we note that the mass capture rate by the neutron star
is much smaller and more strongly phase-dependent for $\alpha_{\rm
  SS}=0.1$ than for $\alpha_{\rm SS}=1$. For $\alpha_{\rm SS}=0.1$,
the phase at which the neutron star captures the disc mass most
slightly lags behind the periastron passage, because of the presence
of a gap between the disc outer radius and the periastron. In the high
resolution simulation, the peak mass capture rate for $42 \le t \le
47$ is $\sim 10^{-10} \rho_{-11} M_\odot{\rm yr}^{-1}$, which is one
order of magnitude smaller than that for $\alpha_{\rm SS}=1$.  The
mass capture rate then decreases by two orders of magnitude by the
apastron passage.  Even in the normal resolution simulation, in which
the disc density around the truncation radius is significantly higher
than that in the high-resolution simulation, the peak mass capture
rate is about a factor of three smaller than that for $\alpha_{\rm
  SS}=1$, and decreases by a factor of fifty by apastron.
Note that similar, strongly phase-dependent accretion was found in
simulations by \citet{al96b} for circumbinary discs around eccentric
binaries.

In the high-resolution simulation with $\alpha_{\rm SS}=0.1$, the
neutron star, on average, captures the disc mass at the rate of $2.3
\times 10^{-11} \rho_{-11} M_\odot{\rm yr}^{-1}$ for $42 \le t \le
47$, while the mean mass-loss rate from the Be star for the same
period is $1.6 \times 10^{-10} \rho_{-11} M_\odot{\rm yr}^{-1}$. This
indicates that, even after three years of disc growth, about 6/7 of
the gas lost from the star still accumulates in the disc, making
the disc continually denser.

The transient nature in the X-ray activity of Be/X-ray binaries
is considered to result from the interactions between the accreted
material and the rotating magnetised neutron star \citep{swr86}. If
the accreted material is dense enough to make the magnetospheric
radius smaller than the corotation radius, the accretion onto the
neutron star causes a bright X-ray emission (the direct accretion
regime). Otherwise, the magnetospheric radius is larger than the
corotation radius, and the accretion onto the neutron star is prevented
by the propeller mechanism. The system is then in quiescence (the
propeller regime).

Recently, \citet{cam01} found that 4U\,0115+63 showed a dramatic X-ray
luminosity variation from $L_{\rm X} \sim 2 \times 10^{34} {\rm
  erg\,s}^{-1}$ to $L_{\rm X} \sim 5 \times 10^{36} {\rm erg\,s}^{-1}$
in less than 15\,hr close to periastron, whereas it showed a nearly
constant X-ray luminosity at a level of $L_{\rm X} \sim 2 \times
10^{33} {\rm erg\,s}^{-1}$ near apastron. They concluded that the
system was in the transition regime between the direct accretion
regime and the propeller regime close to periastron and in the
propeller regime near apastron, because the direct accretion regime
and the propeller regime, respectively, apply for $L_{\rm X} \ga
\xi^{7/2} 10^{37} {\rm erg\,s}^{-1}$ and $L_{\rm X} \la \xi^{7/2} 2
\times 10^{34} {\rm erg\,s}^{-1}$, where $0.5 \le \xi \le 1$ is a
parameter which depends on the geometry and physics of accretion.

Below we try to compare our simulation results with the above criteria
by \citet{cam01}. Although it is likely that the captured material
will form an accretion disc around the neutron star, we do not know
how the accretion rate is related to the mass capture rate shown in
Figs.~\ref{fig:mdot_a1} and \ref{fig:mdot_a01}, which is based on the
moments at which particles enter the variable accretion radius of the
neutron star.  Therefore, we consider two extreme situations, in which
$t_{\rm acc} \ll P_{\rm orb}$ or $t_{\rm acc} \sim P_{\rm orb}$, where
$t_{\rm acc}$ is the accretion time-scale. In the former situation,
the accretion rate profile is approximately the same as the profile of
mass capture rate, whereas in the latter situation, the variation in
the accretion rate will be much smaller than that in the mass capture
rate. We assume that the X-ray luminosity is given by $L_X = \eta GM_X
\dot{M}/R_X$ with $\eta=1$.

For $\alpha_{\rm SS}=1$, $L_{\rm X} \sim 10^{37} \rho_{-11}
M_{\sun}\,{\rm yr}^{-1}$ at periastron if $t_{\rm acc} \ll P_{\rm
  orb}$ and all the captured mass accretes onto the neutron star. Note
that this level of luminosity enters the direct accretion regime.
Since 4U\,0115+63 is considered to be in the direct accretion regime
only in occasional giant X-ray outbursts, we conclude that the
$\alpha_{\rm SS}=1$ model is inconsistent with the observation if
$t_{\rm acc} \ll P_{\rm orb}$. If $t_{\rm acc} \sim P_{\rm orb}$, the
neutron star will emit the X-ray luminosity corresponding to the mean
mass capture rate of $\sim 2 \times 10^{-10} \rho_{-11} M_{\sun}\,{\rm
  yr}^{-1}$ (see the bottom panel of Fig.~\ref{fig:mdot_a1}). The
X-ray luminosity is then $L_{\rm X} \sim 2 \times 10^{36} \rho_{-11}
{\rm erg\,s}^{-1}$. This level of luminosity enters the transition
regime. Hence, the $\alpha_{\rm SS}=1$ model is not ruled out by the
observed constraint if $t_{\rm acc} \sim P_{\rm orb}$.

We have the same conclusion for the $\alpha_{\rm SS}=0.3$ model.
Since it gives the mass capture rate about a half of that for
$\alpha_{\rm SS}=1$, the model is inconsistent with the observation if
$t_{\rm acc} \ll P_{\rm orb}$, but is not ruled out if $t_{\rm acc}
\sim P_{\rm orb}$.

On the other hand, for $\alpha_{\rm SS}=0.1$, $L_{\rm X} \sim 10^{36}
\rho_{-11} M_{\sun}\,{\rm yr}^{-1}$ at periastron even if $t_{\rm acc}
\ll P_{\rm orb}$ and all the captured material accretes onto the
neutron star.  This suggests that the system is in the transition
regime even at periastron. It is likely that the low mass capture
rate in this simulation put the system into the propeller regime in
most of the orbital phases. Thus, the $\alpha_{\rm SS}=0.1$ model for
4U\,0115+63 is consistent with the observed X-ray behaviour,
irrespective of the accretion time-scale.

\section{Summary and Discussion}
\label{sec:summary}

In this paper, we have presented results from 3D SPH simulations of
the disc formation around isolated Be stars and of the interaction
between the Be star disc and the neutron star in Be/X-ray binaries,
based on the viscous decretion disc model for Be stars \citep{lso91}.
In several simulations we adopted constant values of artificial
viscosity parameters $\alpha_{\rm SPH}$ and $\beta_{\rm SPH}$, for
which the Shakura-Sunyaev viscosity parameter $\alpha_{\rm SS}$ is
variable in time and space. In the other simulations, we adopted
constant values of $\alpha_{\rm SS}$, for which $\alpha_{\rm SPH}$ is
variable in time and space and $\beta_{\rm SPH}=0$. We preferred
constant $\alpha_{\rm SS}$ simulations because the analysis of the
results becomes easier.  In all simulations, the Be disc was nearly
Keplerian and the radial velocity was highly subsonic. These features
are consistent with those observed for Be discs.

The simulations of isolated Be stars showed that our code is capable
of producing results similar to those from the 1D simulations. The
simulated mass-loss rate from the Be star in the first several years
of disc formation was several $\times 10^{-10} \rho_{-11} M_{\sun}\,
{\rm yr}^{-1}$ for a wide range of viscosity parameter, which is
consistent with the observed equatorial mass-loss rate.  Here,
$\rho_{-11}$ is the highest local density at $t= 1$\,yr normalised by
$10^{-11}{\rm g}\,{\rm cm}^{-3}$, a typical value for Be stars.

In binary simulations, we have studied the effect of viscosity on the
star-disc interaction in the case of a coplanar system with $P_{\rm
  orb}=24.3\,{\rm d}$ and $e=0.34$, the parameters for 4U\,0115+63,
one of the best studied Be/X-ray binaries. We have chosen these
orbital parameters because they enable us to compare the simulation
results with the observed ones and the short orbital period makes the
computing time bearable.  Some of the results from these simulations
are summarised in Table~\ref{tab:summary}.

Our simulations showed that there is a radius outside which the
azimuthally averaged surface density decreases steeply. For a smaller
$\alpha_{\rm SS}$, the slope outside this radius is steeper, giving
a stronger truncation effect on the disc. Among the simulations with
$\alpha_{\rm SS}=1$, 0.3, and 0.1, we found that the resonant
truncation of the Be disc is effective only for $\alpha_{\rm SS}=0.1$.
For $\alpha_{\rm SS}=1$ and 0.3, the neutron star captures the disc
mass at a rate comparable to the mass-loss rate from the Be star.
These results confirm the previous semi-analytical result by
\citet{no01} and \citet{on01a} on disc truncation that the resonant
truncation is effective for a disc with $\alpha_{\rm SS} \ll 1$. The
truncation radius for $\alpha_{\rm SS}=0.1$ roughly agrees with that
derived semi-analytically.

Our simulations, in particular the high-resolution simulation with
$\alpha_{\rm SS}=0.1$, showed how the disc grows under the influence
of the neutron star.  In the initial build-up phase, the disc
evolution is similar to that for isolated Be stars. But, later on, the
effect of the resonant torque becomes apparent, preventing the disc
gas from drifting outwards at several resonance radii. The effect is
most remarkable at the 4:1 and 5:1 radii ($r/a \sim 0.39$ and 0.33,
respectively) for $\alpha_{\rm SS}=0.1$. As a result, the disc density
increases more rapidly than that for isolated Be stars. This feature
is consistent with \citet{zam01}, who found that the discs in Be/X-ray
binaries are about twice as dense as those of isolated Be stars.

Since the neutron star orbits in an eccentric orbit about the Be star,
the interaction is phase dependent. The disc shrinks a little at
periastron and then recovers gradually. Consequently, the disc
emission will vary with phase. For our system with $e=0.34$, this
effect turned out not to be large enough to cause an observable
variation in the disc continuum as long as it is optically thin. It is
possible, however, that the orbital modulation in the disc continuum
in systems with much higher orbital eccentricity is observable.
Moreover, it is likely that the profiles of optically thick emission
lines from the disc, such as Balmer lines, will show an orbital
modulation even in systems with moderate eccentricity, because the
disc structure is made non-axisymmetric by the periodic excitation and
damping of the spiral density wave.

In the Be disc in Be/X-ray binaries, an eccentric mode is excited
through direct driving due to the (1,0) harmonic of the binary
potential. In high-viscosity simulations, the mode grows initially
linearly in time and then approaches a steady state, of which
steady eccentric shapes based on linear theory of the evolution of a
3D eccentric decretion disc described in Section~\ref{sec:ecc_mode}
offer a fair explanation. The strength of the mode in the steady
state is larger for a smaller value of $\alpha_{\rm SS}$.

On the other hand, in the high-resolution simulation with
$\alpha_{\rm SS}=0.1$, the eccentric mode undergoes prograde
precession at some point. The precession period is about $20P_{\rm
  orb}$. The precession rate is not constant. It accelerates
for $0 < \omega_{\rm d} < \pi$ and decelerates for $\pi <
\omega_{\rm d} < 2\pi$, where $\omega_{\rm d}$ is the angle between
the eccentric vector of the disc and that of the binary orbit. The
precession rate is radius-dependent. It is larger at a
larger radius. Since the eccentric mode is leading for $0 <
\omega_{\rm d} < \pi$ and trailing for $\pi < \omega_{\rm d} <
2\pi$, the twist of the mode increases with time for
$0 < \omega_{\rm d} < \pi$ and decreases for $\pi < \omega_{\rm d} <
2\pi$. The strength of the eccentric
mode, which anti-correlates with the twist of the
mode, decreases for $0 < \omega_{\rm d} < \pi$ and increases for $\pi
< \omega_{\rm d} < 2\pi$.

As the mode precesses, the mass capture rate, $\dot{M}_{\rm acc}$, by
the neutron star modulates.  It is a maximum at $\omega_{\rm d} \sim
\pi$. Even for an eccentricity of 0.1, $\dot{M}_{\rm acc}$ at maximum
is about a factor of three higher than the level before precession. If
the disc had a much stronger disturbance, the enhancement in
$\dot{M}_{\rm acc}$ is likely to be much larger. Such a system may
temporarily show periodic X-ray outbursts.

In our model for Be/X-ray binaries, in which the disc material
overflows toward the neutron star via the $L_1$ point, the mass
capture rate by the neutron star becomes strongly phase dependent. The
dependence is stronger for a smaller value of viscosity. In the disc
with $\alpha_{\rm SS}=0.1$, the mass capture rate decreases by two
orders of magnitude between periastron and apastron passages, and
after apastron passage, no disc mass is captured by the neutron star.
Note that our model gives a much stronger contrast in the mass capture
rate than that expected for the stellar wind accretion.

We have also compared the simulated mass capture rate with the
observed X-ray behaviour of 4U\,0115+63, considering two extreme
situations, in which $t_{\rm acc} \ll P_{\rm orb}$ or $t_{\rm acc}
\sim P_{\rm orb}$, where $t_{\rm acc}$ is the accretion time-scale. We
found that the disc model for $\alpha_{\rm SS}=0.1$ gives a result
consistent with the observation for both situations, whereas the
higher viscosity models for $\alpha_{\rm SS}=1$ and 0.3 are ruled out
unless $t_{\rm acc} \sim P_{\rm orb}$.

Analysing multi-wavelength long-term monitoring observations of
4U\,0115+63, \cite{neg01} found that the Be star undergoes
quasi-cyclic ($\sim 3-5\,{\rm yr}$) activity, losing and reforming its
circumstellar disc. They also found that, at some point, the growing
disc becomes unstable to warping and then tilts and starts precessing.
Type II X-ray outbursts take place after the warping episode. As shown
in this paper, our $\alpha_{\rm SS}=0.1$ model explains many of the
observed features of 4U\,0115+63 in the phase before the warping
occurs. Our simulation, however, showed no dynamical instability, and
therefore is incapable of explaining the warping episode. Including
the effect of radiation from the Be star may turn out to be essential
to have a model that explains the whole cycle of the disc evolution,
which is beyond the scope of this paper.

In this paper, we have concentrated our study on the viscous effects
on star-disc interaction in a coplanar system with fixed orbital
period and eccentricity. This was done not only as a first step to
have a comprehensive understanding of the interaction between the Be
disc and the neutron star in Be/X-ray binaries, but also to have an
archetypal model with which we can compare the results from our future
simulations. In the next paper, we will study the effects of the
misalignment angles between the Be disc and the binary. We will
discuss the effects of the orbital eccentricity in the third paper,
and the fourth paper will conclude the series by studying the effects
of the orbital period.

\section*{acknowledgements}

We thank the anonymous referee for constructive comments. ATO
acklowledges the Institute of Astronomy, Cambridge, UK for the warm
hospitality.  The high-resolution simulation reported here was
performed using the UK Astrophysical Fluids Facility (UKAFF). The
other simulations were done using the facility at the Hokkaido
University Computing Center, Japan. This work was supported in part by
Grant-in-Aid for Scientific Research (13640244) of Japan Society for
the Promotion of Science.

\end{document}